\documentclass[aps,pre]{revtex4}

\usepackage{hyperref}
\usepackage{amsmath}
\renewcommand{\|}{\; | \;}

\newcommand{\BEQ}{\begin{eqnarray}}
\newcommand{\EEQ}{\end{eqnarray}}
\newcommand{\BEA}{\begin{eqnarray}}
\newcommand{\EEA}{\end{eqnarray}}

\newcommand{\hH}{{\hat H}}
\newcommand{\hS}{{\hat S}}
\newcommand{\hSig}{{\hat \Sigma	}}
\newcommand{\hPi}{{\hat \Pi}}
\renewcommand{\d}{{\rm d}}
\newcommand{\half}{{\frac{1}{2}}}
\newcommand{\sop}{{\hSig}}
\newcommand{\s}{{\Sigma}}
\newcommand{\N}{\ensuremath{\mathcal{N}}}
\newcommand{\sigmaN}{{\tilde\sigma}}
\begin{document}
\title{Quantum spherical spin models}
\author{R. Serral Graci\`a}
\author{Th.M. Nieuwenhuizen}
\affiliation{Instituut voor Theoretische Fysica\\
Universiteit van Amsterdam\\ Valckenierstraat 65\\
1018 XE Amsterdam (The Netherlands)}
\email[E-mail:]{rubeng@science.uva.nl, 
nieuwenh@science.uva.nl}

\begin{abstract}
A recently introduced class of quantum spherical spin 
models is considered in detail. Since the spherical 
constraint already contains a kinetic part, the 
Hamiltonian need not have kinetic term.  As a consequence, 
situations with or without momenta in the Hamiltonian can 
be described, which may lead to different symmetry 
classes. Two models that show this difference are 
analyzed.  Both models are exactly solvable and their 
phase diagram is analyzed. A transversal external field 
leads to a phase transition line that ends in a quantum 
critical point. The two considered symmetries of the 
Hamiltonian considered give different critical phenomena 
in the quantum critical region. The model with momenta is 
argued to be analog to the large-$\N$ limit of an SU($\N$) 
Heisenberg ferromagnet, and the model without momenta 
shares the critical phenomena of an SU($\N$) Heisenberg 
antiferromagnet.
\end{abstract}
\maketitle

\section{Introduction} \label{intro}

The classical spherical model was conceived by Kac. After 
being introduced in 1947 to Onsager's rather intricate 
solution of the 2d-Ising model, he desired to formulate a 
simpler spin model. As a first step he took the spins to 
be continuous Gaussian variables, nowadays called the 
Gaussian model. This had unphysical behavior at low 
temperatures which led Kac to consider the ``spherical 
model''. The spherical model has continuous spins that are 
restricted by the ``spherical'' constraint $\sum_{i=1}^N 
S_i^2=N$, which represents the hypersphere intersecting 
all vertices of the hypercube sustained by the Ising 
spins, $S_i=\pm1$. In the end, the spherical model is 
formally the same as the Ising model with a global 
constraint instead of a local one: the sum of spins is 
constrained instead of each of them. At that time the 
saddle point method, needed in the solution, was not 
widely known, and here Berlin came in, leading the 
celebrated joint publication on the spherical model in 
1952 ~\cite{berlin-1952_1}. Kac's personal reminiscence of 
this history is presented in Ref. ~\cite{Kac64}.

The spherical model for a ferromagnet has been considered 
in great detail. Actually, the paramagnetic to 
ferromagnetic transition is similar to an ideal 
Bose-Einstein condensation. Since the solution of the 
model is so simple and explicit, the critical behavior can 
be solved exactly. Critical exponents and scaling 
functions can be derived. In particular, the model with 
short range interactions exhibits $d_{\rm lc}=2$ as the 
lower critical dimension; for $d\le 2$ no stable 
ferromagnetic phase occurs. Likewise, $d_{\rm uc} =4$ is 
the upper critical dimension; for $d>4$ critical exponents 
take their mean-field values. These analytic results have 
been used to test approximations and general ideas of 
phase transitions for a wide range of interactions, short 
and long range. For a review on the classical spherical 
model see Ref.~\cite{joyce}.

As said, the spherical model was introduced for its 
mathematical simplicity. However, Stanley 
\cite{stanley-1968_1} proved that the free energy of a 
model of arbitrary spin dimension $\nu$, incorporating 
thus the Ising model (for spin dimension $\nu=1$), the 
$x-y$ model (spin dimension $\nu=2$) and Heisenberg model 
($\nu=3$), approaches that of the spherical model in the 
limit of infinite spin dimensionality $\nu \to \infty$. 
Hence, it gives a geometrical interpretation to the 
spherical model. Since various critical properties where 
proven to be monotonic functions of the spin 
dimensionality $\nu$, the critical properties of the 
Heisenberg model appeared to be bounded on one side by 
those of the Ising model and on the other by those of the 
spherical model.

The spherical model for antiferromagnets was thoroughly 
studied by Knops. The spherical constraint imposes 
$<S^2_i>=1$ for ferromagnets. However, this does not work 
for antiferromagnets because of the lack of translational 
invariance. To recover this, he added a second constraint; 
more generally, one constraint has to be added for each 
translationally invariant set, which in the case of 
antiferromagnets means each of the two sublattices. Knops 
found that the two constraints reduce to a unique one 
provided the staggered external field vanishes. The fact 
that the spherical spins are scalars make it impossible to 
define an order parameter that can be identified with the 
spontaneous staggered magnetization. To solve that and get 
the proper order parameter Knops used a vector version of 
the spherical model \cite{knops-1973_2}. He also 
generalized Stanley's arguments to non-translational 
interactions ~\cite{knops-1973_1}.

The spherical model has also been applied to disordered 
systems. Though, in view of Knops' finding, perhaps an 
infinite number of spherical constraints should be used, 
typically no analog of the staggered external field is 
applied, and one may expect that all constraints collapse 
into a single one. Therefore  spherical spin glass models 
may still give insight in the physics of the problem which 
would be more difficult to study for e.g. Ising spins. In 
the case of pair couplings the exact solution exhibits no 
breaking of replica symmetry and the replica trick need 
not be used ~\cite{kosterlitz-1976_1}. The family of 
$p$-spin spin glasses ($p$-spin models) 
~\cite{crisanti-1992_1}  has been shown to exhibit one 
step replica symmetry breaking by studying the spherical 
version. For spin glasses with random pair and quartet 
interactions ($\{p=2\}+\{p=4\}$, $``p=2+4''$), one of us 
showed that an exact solution exists, exposing the full 
replica symmetry breaking scenario. The simplicity of 
spherical models thus may give insight in difficult 
problems for which otherwise no exact solution is 
available. For an early review on the use of the spherical 
model in disordered systems, see Ref.~\cite{khorunzhy}.

So far the discussion has been classical. The classicality 
can be understood in particular because the entropy 
diverges at low temperature as $\ln T$, just as for a 
classical ideal gas. Different quantum versions of the 
spherical model have been proposed. Obermair studied 
surface effects in phase transitions ~\cite{obermair}. 
Identifying with a spin an operator $\hat S_i$, he 
postulated a momentum operator  $\hat\Pi_i$  conjugate to 
it, $[\hat S_i,\hat \Pi_j]=i\hbar\delta_{ij}$. To get a 
spin dynamics, he added a kinetic term $\half g \sum_i 
\hat\Pi_i^2$ to the Hamiltonian, but kept the constraint 
the same except for expressing it in operators as 
$\Sigma_i\langle\hat S_i^2\rangle=N$. In this case, the 
kinetic term may be understood as the kinetic energy of 
rigid rotors. The model remains exactly solvable. Many 
others have therefore used this quantization in the study 
of spin glasses \cite{shukla-1981_1}, systems with 
multispin random interactions (p-spin glasses) 
\cite{cugliandolo-2000_1}, or the study of quantum phase 
transitions \cite{vojta-1996_1}. 

One of us presented in 1995 a different quantum approach 
to cure the low temperature behavior 
\cite{nieuwenhuizen-1995_2}. In a Trotter approach to the 
partition sum, the first step is to take as spherical 
constraint: $\sum_i \Sigma^\ast_i 
\Sigma_i=N\sigmaN/\hbar^2$, where $\sigmaN$ is a constant 
that need not be unity, and $\Sigma_i$ is the complex 
parameter characterizing the coherent state associated 
with the bosonic annihilation operator $\hat\Sigma_i=[\hat 
S_i/\hbar+i\hat \Pi_i]/\sqrt{2}$. Hence, in this approach 
the momentum appears in the constraint. Indeed, this 
constraint may also be written $\sum_{i}\langle 
\hS_i^{2}\rangle +\hbar^2\langle \hPi_i^{2}\rangle 
=2N\sigmaN$. As a second step, momenta dependent 
Hamiltonians were considered, by replacing 
$J_{ij}S_iS_j\to J_{ij} \hat \Sigma_i^\dagger 
\hat\Sigma_j$. Later \cite{nieuwenhuizen-1998_1}, the same 
formalism was applied to the p-spin glass model and was 
compared with its Ising counterpart. In spite of the 
simplicity of the system and its solubility, the resulting 
phase diagram shows very interesting critical phenomena. 
Since the momenta are present in the constraint one may 
also study situations where they does not appear in the 
Hamiltonian. 

Below we will consider two Hamiltonians with nearest 
neighbor ferromagnetic interactions. We will see that the 
presence or absence of momenta can change the symmetries 
of the action giving rise to different critical phenomena 
in the quantum region. The resulting action may be 
invariant under unitary transformations or under 
orthogonal ones, while in Obermair's approach only the 
latter is possible. In the last section we show that one 
of these spherical spin models relates to a quantum 
ferromagnet and the other to a quantum antiferromagnet.

The two different quantum versions of the model have the 
same quantization rule $[\hat 
S_i,\hat\Pi_j]=i\hbar\delta_{i,j}$. Both of them cure the 
problem of the entropy, it remains positive and, for 
temperature going to zero, goes to zero as a power law. 
Vojta, \cite{vojta-1996_1} following Stanley's arguments, 
found that Obermair's quantization gives a free energy 
that is identical to the large-$n$ limit of the O($n$) 
nonlinear sigma model. Therefore it describes rotors 
instead of Heisenberg spins. Nieuwenhuizen, conversely, 
gave indications that his version had a behavior closer to 
Heisenberg spins, as having  in the case of free spins a 
gap between the ground and the first excited state scaling 
with the field at small fields. 

The aim of the present paper is to point out the 
differences between these two models. We study two 
Hamiltonians using Nieuwenhuizen's spherical constraint. 
The first one was introduced in 
Ref.~\cite{nieuwenhuizen-1995_2} and we find it to be 
analogous to the large-$\N$ limit of a SU($\N$) Heisenberg 
ferromagnet. In the second Hamiltonian studied, no momenta 
are present; momenta only appear in the formalism through 
the spherical constraint. In this case, the same critical 
phenomena as in the large-$\N$ limit of a SU($\N$) 
Heisenberg antiferromagnet is found, which can be 
described by an O($\N$) nonlinear $\sigma$-model which, in 
turn, is analogous to Obermair's model.

The paper is organized as follows: in section 
\ref{classical} the classical spherical model is reviewed 
and the way to quantize it is discussed. The differences 
between the two quantum spherical constraints are pointed 
out. In section \ref{pathint}, the path integral formalism 
is introduced to calculate the partition function of a 
quantum spherical model. In section \ref{complex} this 
formalism is used to solve the thermodynamics of a 
ferromagnetic quantum spherical model with nearest 
neighbor interaction and the critical phenomena is studied 
in detail. At the finite temperature phase transition, the 
critical exponents remain the same as the classical ones. 
In section \ref{realpart} a Hamiltonian with the same 
couplings but without momenta is considered. The critical 
exponents are found to be the same as in Obermair's model. 
After that, in section \ref{mapping} a generalization of 
the two types of Hamiltonians presented here is given and 
the limit of SU($\N$) Heisenberg spins is argued to give 
the same critical behavior as this quantum version of 
spherical model. Finally some conclusions are drawn.

\section{Classical spherical model} \label{classical}

The spherical constraint was conceived as a relaxation of 
the Ising constraint. Indeed, Ising spins, $S_i\equiv 
s_{i,z}=\pm \half\hbar$, obviously satisfy it. Adjusting 
the coefficients from the original version it may be 
written as \BEQ \label{concl} \half \sum_{i=1}^N 
S_i^2=N\sigma, \EEQ with $\sigma={\hbar^2}/{8}$ having 
dimension $(\text{Js})^2$. The Berlin-Kac spherical model 
is defined by the partition sum \BEQ Z=\int DS\,e^{-\beta 
H}\,\delta(\half\sum_{i=1}^N S_i^2-N\sigma) =\int 
DS\,\int_{-i\infty}^{i\infty} \frac{\d\tilde\mu}{2\pi 
i}e^{-\beta H- \half \tilde\mu\sum_{i=1}^N S_i^2 
+\tilde\mu N\sigma}
\label{eq:bkz}
\EEQ
where 
\begin{eqnarray}
DS=\prod_i \int_{-\infty}^\infty \d S_i
\end{eqnarray}

\subsection{Vector spherical spins}
For vector spins the generalization of eq. (\ref{concl}) 
in the case of $m$ spin dimensions reads

\begin{equation}
\half \sum_{i=1}^N \sum_{a=1}^m {(S_i^a)}^2 = N m \sigma
\label{eq:veccon}
\end{equation}

It is worth mentioning that the spin dimensionality in eq. 
(\ref{eq:veccon}) is not related to the approach of 
Stanley, who started with vector spins and ended up with 
scalar spherical spins. We only introduce vector spherical 
spins to avoid the restriction scalar spins have. We 
benefit from the fact that the vector character allows to 
study the behavior in a transverse field. A similar step 
allowed Knops to define a proper order parameter for the 
antiferromagnetic spherical model ~\cite{knops-1973_2}.  

\subsection{Quantization}
It is natural to consider the $S_i$ analogous to position 
variables of harmonic oscillators. In quantum mechanics 
they become hermitian operators $\hat S_i$ with the 
dimension of $\hbar$, Js. The conjugate momentum operator  
$\hat\Pi_i^a$ is dimensionless and postulated to satisfy 
the commutation relation
\BEQ [\hS_i^a,\hat \Pi_j^b]=i\hbar\delta_{i,j}\delta_{a,b} 
\EEQ
As for harmonic oscillators, this allows to define 
creation and annihilation operators
\BEQ 
\hSig_i^{a\,\dagger}=\frac{1}{\hbar\sqrt{2}}\hS_i^a-
\frac{i}{\sqrt{2}}\hPi_i^a,\qquad 
\hSig_i^a=\frac{1}{\hbar\sqrt{2}}\hS_i^a+\frac{i}{\sqrt{2}}
\hPi_i^a \label{eq:cran}\EEQ
satisfying the commutation relation
\BEQ [\hSig_i^a,\hSig_j^{b\,\dagger}]=\delta_{i,j}
\delta_{a,b} \EEQ

\subsection{Spherical constraint on the length of the 
total spin}

There is some freedom to choose the spherical constraint, 
which amounts to describing different physical situations. 
The standard quantum constraint considered in literature 
is just the quantized version of the mean of eq. 
(\ref{eq:veccon}),

\BEQ \label{eq:const1}
{\rm Constraint\,1:}\qquad \half \sum_{i,a} <(\hS_i^a)^2>
=Nm\sigma \EEQ
where $<...>$ denotes the quantum expectation value.
Obermair took as the quantum Hamiltonian the classical 
$H({\bf S})$ with spins replaced by operators, and added 
the kinetic term that one expects for physical rotors,
\BEQ 
\hH({\bf \hS},{\bf \hPi})=\frac{1}{2}g\sum_i\hPi_i^2+
H({\bf \hS})
\EEQ
where $g^{-1}$ is the rotor's moment of inertia.
An effective Hamiltonian which includes the constraint can 
be derived with a Lagrange multiplier. One ends up with

\BEQ 
\hH_{\rm tot}= \frac{1}{2}g\sum_i\hPi_i^2+H({\bf \hS})
+\mu(t)[\half \sum_{i,a}(\hS_i^a)^2-Nm\sigma]
\EEQ
where $\mu$ is the 
Lagrange multiplier 
that enforces the constraint. In equilibrium its value is
given by the equation of the spherical constraint 
$\frac{\partial F}{\partial \mu}=0$. 
The dynamics is now fixed by the Heisenberg equations of 
motion,
\BEA \label{dyn1}
\frac{\d \hat S_i^a}{\d t}&=&i[\hat H_{\rm tot}(t),\hat 
S_i^a(t)]= g\hat \Pi_i^a(t),\\
\frac{\d \hat \Pi_i^a}{\d t}&=&i[\hat H_{\rm tot}(t),
\hat \Pi_i^a(t)] = -\frac{\partial \hat H}{\partial 
\hat S_i^a} -\mu(t)\hat S_i^a(t).\label{dyn2} \EEA
where the real parameter $\mu$ has to be taken 
time-dependent in order to satisfy the ``soft'' constraint 
(\ref{eq:const1}) at each instant. It is clear that a 
non-zero $g$ is needed to get any spin dynamics. Combining 
the two equations one has
\BEA 
\frac{1}{g}\,\frac{\d^2 \hS_i^a}{\d t^2}
= -\frac{\partial \hat H}{\partial \hat S_i^a}
- \mu(t)\hat S_i^a(t).\label{dyn2a} \EEA

Is worth remarking that no energy budget is involved in 
the spherical constraint,
\BEQ \langle\hat H_{\rm tot}\rangle =\langle\hat H\rangle.
\EEQ

\subsection{Spherical constraint on the number of spin 
quanta}

In 1995 one of us has proposed a constraint that fixes the 
number of quanta ~\cite{nieuwenhuizen-1995_2}. In a path 
integral approach it was assumed that the $c$-numbers 
$\s_i^a$, which characterize a coherent state, satisfy at 
each timestep
 \BEQ {\rm Constraint\,2':}\qquad 
\sum_{i,a}\s_i^{a\,\ast}\s_i^a=Nm\frac{\sigmaN}{\hbar^2}.
\label{eq:const2a}
\EEQ
It is to be expected that this is equivalent to
\BEQ {\rm Constraint\,2:}\qquad 
\sum_{i,a}<\hSig_i^{a\,\dagger}\hSig_i^a>=\sum_{i,a}<\hat 
n_i^a>= Nm\frac{\sigma}{\hbar^2}.
\label{eq:const2}
\EEQ

We shall show below that this is indeed the case, and the 
relation, $\sigmaN=\sigma+\hbar^2$, is derived in Eq. 
(\ref{sisiN=}). This constraint includes the momenta as 
can be seen by writing it in the form
\BEQ {\rm Constraint\,2:}\qquad 
\half\sum_{i,a}(\langle \hS_i^{a\,2}\rangle
+\hbar^2\langle \hPi_i^{a\,2}\rangle)
=Nm(\sigma + \frac{\hbar^2}{2}) \EEQ
For a Hamiltonian $\hH({\bf \hS},{\bf \hPi})$ that may, 
but need not, depend explicitly on the momenta, the 
effective spherical Hamiltonian is
\BEQ 
\hH_{\rm tot}=\hH({\bf \hS},{\bf \hPi})+
\half\mu\sum_{i,a}\left[(\hS_i^a)^2+\hbar^2(\hPi_i^a)^2
\right] -Nm\mu(\sigma+\frac{\hbar^2}{2}) 
\label{eq:htot}
\EEQ

Now, the situation where the Hamiltonian does not depend 
explicitly on the momenta (no kinetic term), $\hH({\bf 
\hS},{\bf \hPi})\to\hat H({\bf \hat S})$, still leads to 
sensible dynamics, since the constraint already depends on 
the momenta. Different constraints describe different 
physics. However, at high temperatures one expects the 
differences to become small.

Eq. (\ref{dyn1}) now brings $\hat\Pi_i^a=(\d \hat S_i^a/\d 
t)/\mu(t)$, eq. (\ref{dyn2}) remains the same. They may be 
combined together into a second order equation for the 
spin operators,
\BEA \label{dyn3}
\frac{\d}{\d t}\left(\frac{1}{\mu(t)}\,\frac{\d \hat 
S_i^a}{\d t}\right) &=&-\frac{\partial \hat H}{\partial 
\hat S_i^a} -\mu(t)\hat S_i^a(t). 
\EEA

In the remaining of this paper we will simplify the 
notation by taking units in which $\hbar=1$.

\subsection{Comparison of the two constraints}

The main difference between the two constraints is 
obviously the presence or absence of momenta. In the 
second case eq. (\ref{eq:const2}), the spherical 
constraint can carry all the dynamics of the model. On the 
contrary, using the first constraint eq. 
(\ref{eq:const1}), a kinetic term, with an external 
parameter $g$, has to be added to the Hamiltonian 
~\cite{obermair}. This parameter determines the strength 
of quantum fluctuations; the classical model can be 
recovered for $g=0$. This fact makes models with the first 
constraint describe quantum rotors, as was pointed out in 
Ref.~\cite{vojta-1996_1}. The first constraint, eq. 
(\ref{eq:const1}), brings actions which are invariant 
under orthogonal transformations. Conversely, using the 
second constraint, eq. (\ref{eq:const2}), the choice of 
Hamiltonian can bring symmetry under unitary 
transformations or orthogonal ones depending on the 
question whether the Hamiltonian contains momenta or not. 
Hamiltonians with unitary transformation symmetry yield 
free energies analogous to the large $\N$ limit of the 
generalization of SU(2) Heisenberg spins to SU($\N$). 
Hamiltonians with orthogonal transformation symmetry share 
the critical phenomena with the large $\N$ limit of 
O($\N$) non-linear sigma model and describe therefore 
quantum rotors as occurs by using the first constraint, 
eq. (\ref{eq:const1}).

Each of the symmetries belong in different universality 
classes in the quantum regime, yet classical critical 
phenomena are always the same as in the classical model, 
consistent with the expectation that quantum effects do 
not lead to qualitative changes at finite temperatures. We 
will see that the dynamical critical exponent $z$ is be 
different in both symmetries, causing the difference in 
critical exponents at the quantum critical point as was 
pointed out in Ref. \cite{hertz-1976_1}.

\section{Path Integrals} \label{pathint}
In this section we explain, following Ref. 
~\cite{nieuwenhuizen-1998_1}, how to add the spherical 
constraint to a quantum Hamiltonian using the path 
integral formalism for models with the second constraint 
eq. (\ref{eq:const2}). In second quantization the spins 
are given a bosonic algebra. In the path integral the 
boson coherent state representation is used for the spins 
(for a review of path integrals and coherent states, see 
e.g. ~\cite{negele} and for a complete study of coherent 
states see e.g. ~\cite{klauder}).

\subsection{Bosonic coherent state representation for a 
single oscillator}

Fock space is the Hilbert space of states labeled by the 
number of oscillator quanta. Coherent states are defined 
as the eigenstates of the anihilator operator ${ \hat a}$. 
Then it can be proved that for a system with many particles

\begin{equation}
\|\phi \; \rangle = e^{\sum_\alpha \phi_\alpha { \hat 
a^\dagger_\alpha}} \|0 \; \rangle = \prod_{\alpha} \left\{ 
\sum_{n_{\alpha}} \frac{(\phi_{\alpha} { \hat 
a^\dagger}_{\alpha})^{n_{\alpha}}} {n_{\alpha}!} \right\} 
\|0\; \rangle
\end{equation}

\noindent is a coherent state, where $\|0\; \rangle$ is 
the vacuum representation in Fock's space, and $\alpha$ 
stands for each state for any particle of the system. 
Indeed, because of the identity ${\hat a}_\alpha({\hat 
a}^\dagger_\alpha)^{n_\alpha} = n_\alpha ({\hat 
a}^\dagger_\alpha)^{n_\alpha-1} + ({\hat 
a}^\dagger_\alpha)^{n_\alpha} {\hat a}_\alpha$, it holds 
that ${\hat a}_\alpha \| \phi \rangle = \phi_\alpha 
\|\phi\rangle$. The scalar product of two coherent states 
gives

\begin{equation}
\langle \;\phi \| \phi'\; \rangle = 
e^{\sum_\alpha \phi^\ast_\alpha \phi'_\alpha}
\label{eq:norm}
\end{equation}

A crucial property of the coherent states is that they 
form an overcomplete set of states. Any vector in Fock 
space can then be expanded in terms of coherent states. 
This is expressed by the closure relation \cite{negele} 

\begin{equation}
\int \prod_\alpha \frac{d \Im (\phi_\alpha) d \Re 
(\phi_\alpha)}{\pi} e^{-\sum_\alpha \phi^\ast_\alpha 
\phi_\alpha} \|\phi\; \rangle\langle 
\;\phi\| = {\bf 1}
\label{eq:identity}
\end{equation}
where the measure in the integral comes from gaussian 
integration with complex variables ($\Im$ stands for 
imaginary part and $\Re$ for real part) and the 
exponential term is due to the fact that coherent states 
are not normalized. Let us check that eq. 
(\ref{eq:identity}) is indeed a representation of the 
identity of Fock space. We insert it in the left hand side 
of eq. (\ref{eq:norm}) and we get

\begin{equation}
\begin{split}
\langle \;\phi \| {\bf 1}  \| \phi'\; \rangle =\int 
\prod_\alpha \frac{d \Im (\psi_\alpha) d \Re 
(\psi_\alpha)}{\pi} e^{-\sum_\alpha \psi^\ast_\alpha 
\psi_\alpha} \langle \; \phi \|\psi\; \rangle\langle 
\;\psi\| \phi'\; \rangle = \\
\int \prod_\alpha \frac{d \Im (\psi_\alpha) d \Re 
(\psi_\alpha)}{\pi} e^{\sum_\alpha -(\psi^\ast_\alpha 
\psi_\alpha - \phi^\ast_\alpha \psi_\alpha - 
\psi^\ast_\alpha \phi'_\alpha)} = e^{\sum_\alpha 
\phi^\ast_\alpha \phi'_\alpha}
\end{split}
\end{equation}
which indeed is the right hand side of eq. (\ref{eq:norm}).

The partition function of any quantum system $Z={\rm 
tr}\,\, e^{-\beta H({ \hat a^\dagger},{ \hat a})}$ can be 
computed by the Trotter approach. The exponential has the 
same form as a time evolution operator in imaginary time. 
Thus it is possible to create a path integral over closed 
paths.  The procedure is to split the exponential in a 
product of $M$ equal terms. Between each pair of them a 
representation of the identity, eq. (\ref{eq:identity}), 
is inserted. The partition sum then has the following shape

\begin{equation}
Z= {\rm tr} \left\{ \left( e^{-\epsilon H({ \hat 
a^\dagger},{ \hat a})}\right)^M \right\} =
{\rm tr} \; \left\{ e^{-\epsilon H({ \hat a^\dagger},{ 
\hat a})} \, {\bf 1}\, e^{-\epsilon H({ \hat a^\dagger},{ 
\hat a})}{\bf 1} \cdots \, {\bf 1}\, e^{-\epsilon H({ \hat 
a^\dagger},{ \hat a})} \, \right\} 
\end{equation}
where $\epsilon = \beta/M$ and each ${\bf 1}$ is an 
identity operator. Each of these identities is given an 
index; they represent the steps the system passes through 
in a discretized path. By using the identity defined in 
Eq.~(\ref{eq:identity}) the following matrix element is 
needed: 
\begin{equation}
\langle \; {\bf \phi}_{j}\| 
e^{-\epsilon H({ \hat a^\dagger},{ \hat a})} 
\|{\bf \phi}_{j-1} \;\rangle 
\end{equation}
Provided the Hamiltonian is normal ordered, the outcome 
is ~\cite{negele}
\begin{eqnarray}
\langle \; {\bf \phi}_{j}\| e^{-\epsilon H({ \hat 
a^\dagger},{ \hat a})} \|{\bf \phi}_{j-1} \;\rangle 
&\approx& \langle \; {\bf \phi}_{j}\| 1-\epsilon H({ \hat 
a^\dagger},{ \hat a}) \|{\bf \phi}_{j-1} 
\;\rangle = e^{{\bf \phi}_{j}^\ast\cdot{\bf 
\phi}_{j-1}}(1- \epsilon H({\bf \phi}_{j}^\ast,{\bf 
\phi}_{j-1}))
\nonumber \\
&=& e^{{\bf \phi}_{j}^\ast\cdot{\bf \phi}_{j-1}- \epsilon 
H({\bf \phi}_{j}^\ast,{\bf \phi}_{j-1})}
+{\cal O}(\epsilon^2)
\label{eq:path}
\end{eqnarray}

Correction terms can be neglected in the limit 
$M \to \infty$ \cite{negele}. Each identity brings an 
integral at each time step. These integrals cover any 
path between its initial and its final state. The trace 
will finally tie the ends giving a closed path.
The partition function finally reads

\begin{equation}
Z= \int_{\phi_\alpha (\beta)= \phi_\alpha (0)} 
D(\phi^\ast_\alpha (\tau) 
\phi_\alpha (\tau)) \exp \left\{ \sum_{\tau=0}^\beta d\tau 
\left[{\bf \phi}^\ast(\tau)\cdot \frac{d {\bf 
\phi}(\tau)}{d \tau}+ H({\bf \phi}^\ast(\tau),{\bf 
\phi}(\tau-d\tau))\right] \right\}
\end{equation}
where the subindex of the integral reflects the trace 
structure of the partition function since it gives a 
closed path integral; $\tau$ stands for the imaginary time 
step, so $\phi(\tau)=\phi_i$; $d \tau$ is the imaginary 
time difference between steps, so $\phi(\tau - d 
\tau)=\phi_{i-1}$; and 

\begin{equation}
\frac{d \phi(\tau)}{d \tau} = \frac{\phi(\tau)-\phi(\tau - 
d \tau)}{d \tau}= \frac{\phi_i-\phi_{i-1}}{\beta/M}
\end{equation}

Despite the fact that the nomenclature used in these 
formulas suggests a continuous time, it should always be 
understood as being discrete. The limit $M \to \infty$ 
should always be taken at the end of the calculations, 
otherwise some indeterminacies may arise. Continuous 
notation is used nevertheless because it is more compact.

\subsection{Coherent state representation for spherical 
spins}

We can deal with spherical spins using almost the same 
approach. The operator $\hat a_i$ is identified with 
$\sop_i^a$, where the index $a$ denotes the spin vector 
direction, and the corresponding fields $\phi_i$ are 
denoted as $\s_i^a$. We remind that the spherical 
constraint we use is the one defined in eq. 
(\ref{eq:const2}).

In order to impose this constraint in the path integral 
formalism, the identity definition eq. (\ref{eq:identity}) 
is modified to adopt to the spherical case, in a way 
inspired by Ref.~\cite{nieuwenhuizen-1998_1}: one 
restricts the path integral to states which exactly 
satisfy the constraint by employing the truncated identity

\begin{equation}
{\bf 1}\to{\bf 1}_{spherical}
\equiv C\int \prod_{ia}\frac{d \Im (\s_i^{a}) d \Re 
(\s_i^a)}{\pi} e^{-{\bf \s^\ast\cdot \s}} \|{\bf \s}\; 
\rangle \langle \; {\bf \s} \| \delta({\bf \hat n}-Nm\sigma)
\label{eq:i}
\end{equation}
where the number operator ${\bf  \hat n}$, 
\BEQ {\bf  \hat 
n}=\sum_{i,a}\hat\s_i^{a\,\dagger}\hat\s_i^a, \EEQ
counts the total number of spin quanta. We insert 
\begin{equation} \delta({\bf \hat n}-Nm\sigma)
= \int_{-\infty}^{\infty} \frac{\epsilon 
\d \tilde \mu}{2\pi }e^{-i\epsilon \tilde\mu({\bf  \hat 
n}-Nm\sigma)}
= \int_{-i\infty}^{i\infty} \frac{\epsilon 
\d \mu}{2\pi i}e^{-\epsilon \mu({\bf  \hat n}-Nm\sigma)}
\end{equation}
where $\mu =i\tilde \mu$ is imaginary. (Strictly speaking, 
we should insert a Kronecker-$\delta$ function, rather 
than the Dirac-$\delta$, but for
large $N$ this amounts to the same.)
Repeating the same procedure with this new identity we get

\begin{equation}
Z=\int_{\s(\beta)= \s(0)} D\mu D{\bf \s}^\ast D{\bf 
\s}\exp(-{\bf A}),
\label{eq:part1}
\end{equation}
with the action

\begin{equation} \label{action=}
{\bf A}=\sum_{\tau=0}^\beta d\tau \left[{\bf 
\s}^\ast(\tau)\cdot \frac{d 
{\bf \s}(\tau)}{d \tau}+ 
\mu(\tau)({\bf \s}^\ast(\tau)\cdot{\bf \s}(\tau -d\tau
)-Nm\sigma) + H({\bf \s}^\ast(\tau),{\bf 
\s}(\tau-d\tau))\right] 
\end{equation}
and integration measures defined as

\begin{subequations}
\begin{eqnarray}
\int D{\bf \s}^\ast D{\bf \s}&=&\prod_{ia\tau} 
\int_{-\infty}^\infty\int_{-\infty}^\infty \frac{d 
\Im(\s_{i}^a(\tau))d 
\Re (\s_{i}^a(\tau))}{\pi} \label{equationa} \\
\int D\mu &=&C\prod_\tau \int_{-i\infty}^{i\infty} 
\frac{\epsilon 
d\mu(\tau)}{2\pi i} \label{equationb}
\end{eqnarray}
\end{subequations}

\noindent so $\mu(\tau)$ is the Lagrange multiplier 
introduced to impose the spherical constraint and the 
prefactor $C^{(M)}$ is added to ensure, if needed, a 
proper normalization. Details on this factor were given in 
Ref. \cite{nieuwenhuizen-1998_1}.

It should be noted that the particle number operator in 
the definition of the identity eq. (\ref{eq:i}) will be 
surrounded, as is the case for the Hamiltonian, by spin 
operators on different timesteps; therefore its creation 
and annihilation operators will also be projected on 
different timesteps, $\hat\s_i^{a\,\dagger}\hat\s_i^a\to 
\s_i^{a\,\ast}(\tau)\s_i^a(\tau-\d\tau)$. In 
Refs.~\cite{nieuwenhuizen-1995_2,nieuwenhuizen-1998_1} the 
spherical constraint was slightly different from the one 
presented here. The proposal was to take the constraint 
not in terms of the particle number operator but in terms 
of its generating variables (which are $c$-numbers), at 
every imaginary timestep, 
\begin{equation}
{\bf \s^\ast\cdot \s}=\sum_{i=1}^N\sum_{a=1}^m 
\s_i^{a\ast}(\tau) \s_i^a(\tau)=Nm\sigmaN
\label{eq:sphc1}
\end{equation}
so they acquire the same time-index.
The two actions then differ only in the timestep 
projection of the spherical constraint, so with this 
constraint one obtained ${\bf \s}^\ast(\tau)\cdot{\bf 
\s}(\tau)$ rather than ${\bf \s}^\ast(\tau)\cdot{\bf 
\s}(\tau-d\tau)$. The difference that this brings can be 
seen as follows. Starting from eq. (\ref{eq:part1}) we 
want to have two operators projected at the same time. To 
achieve this, it turns out that we must exchange the  
order of the operators, $\s^\dagger \s = \s \s^\dagger - 
1$. The term $\s \s^\dagger$ can be projected at a single 
time as one can see following eq. (\ref{eq:path}). For a 
single component spin at timestep $j$ the relevant matrix 
element is
\begin{gather}\langle \s_{j+1} | e^{-\epsilon \hat H} 
e^{-\epsilon\mu\hat \s^\dagger\hat\s} e^{-\epsilon \hat H} 
|\s_{j-1} \rangle 
= \langle \s_{j+1} | e^{-\epsilon \hat H} 
(1+\epsilon\mu-\epsilon\mu\hat \s\hat \s^\dagger) 
e^{-\epsilon \hat H} |\s_{j-1} \rangle
\notag \\ = (1+\epsilon\mu)\langle \s_{j+1} | 
e^{-\epsilon \hat H} |\s_{j} \rangle \langle \s_{j} |
e^{-\epsilon \hat H} |\s_{j-1} \rangle
-\epsilon\mu \langle \s_{j+1} | e^{-\epsilon \hat H} 
\hat \s |\s_{j} \rangle \langle \s_{j} | \hat \s^\dagger
e^{-\epsilon \hat H} |\s_{j-1} \rangle  \notag \\
= (1+\epsilon\mu-\epsilon\mu \s_j^\ast\s_j)\langle 
\s_{j+1} | e^{-\epsilon \hat H} |\s_{j} \rangle  \langle 
\s_{j} |
e^{-\epsilon \hat H} |\s_{j-1} \rangle 
= e^{\epsilon\mu-\epsilon\mu \s_j^\ast\s_j}\langle 
\s_{j+1} | e^{-\epsilon \hat H} |\s_{j} \rangle \langle 
\s_{j} |
e^{-\epsilon \hat H} |\s_{j-1} \rangle  
\end{gather}
Thus a factor $e^{\epsilon\mu-\epsilon\mu 
\s_{i}^{a\,\ast}(\tau) \s_{i}^a(\tau)}$ comes for each 
spin operator $\hat\s_i^a$ at each timestep. This leads to 
the spherical constraint 
\begin{equation}
{\bf 1}_{spherical}
\equiv C\int \prod_{ia}\frac{d \Im (\s_i^{a}) d \Re 
(\s_i^a)}{\pi} e^{-{\bf \s^\ast\cdot \s}} \|{\bf \s}\; 
\rangle \langle \; {\bf \s} \| 
\delta(\sum_{i,a}\s_i^{a\,\ast}\s_i^a-Nm-Nm\sigma)
\label{eq:ia}
\end{equation}
that with definition eq. (\ref{eq:sphc1}) should be 
compared to eq. (\ref{eq:i}). In words, the spherical
constraint can indeed be taken on the coherent state 
variables as in eq. (\ref{eq:sphc1}) or in 
Ref.~\cite{nieuwenhuizen-1995_2},
provided one makes the identification $\sigmaN=\sigma+1$, 
or, restoring units,
\BEQ \label{sisiN=}\qquad \sigmaN=\sigma+\hbar^2\EEQ
In Eq. (\ref{Theoconn}) we will verify that with this 
identification the two approaches indeed yield the same 
free energy.

It is worth remarking that we imposed the spherical 
constraint strictly, no thermal average has been performed. 
In the following section $\mu$ will be integrated over by 
the method of steepest descends, a procedure that allows 
the particle number to fluctuate; therefore the 
satisfiability of the constraint remains only in average. 

\section{Ferromagnetic Hamiltonians with creation and 
annihilation operators} \label{complex}

Using the formalism described in \ref{pathint} we can 
study the Hamiltonian

\begin{equation}
H({\bf \sop}^\dagger, {\bf \sop}) =  -\sum_{i\neq 
j}J_{ij}{\bf \sop}_{i}^\dagger{\bf \sop}_{j}-\sum_i 
\Gamma_i \frac{( {\bf \sop}^{\dagger}_{i}+{\bf 
\sop}_{i})}{\sqrt{2}} = -\frac{1}{2}\sum_{i\neq 
j}J_{ij}({\bf \hS}_i{\bf \hS}_j + {\bf \hat \Pi}_i {\bf 
\hat \Pi}_j) - \sum_i \Gamma_i {\bf \hS}_i
\label{eq:complexham}
\end{equation}
where in the second equality we inserted eq. 
(\ref{eq:cran}). The $i\hS_i {\hat \Pi}_j$ cancelled since 
we assumed symmetric couplings, $J_{ij}=J_{ji}$. Obviously, 
the momentum operators do occur in this expression. 
The couplings $J_{ij}$ can in principle express any kind 
of interaction, ferromagnetic, antiferromagnetic, spin 
glass... The $\Gamma_i$ represent an external field, that 
can be constant, variable, random... Later on, we will 
focus on ferromagnetic couplings in the presence of 
constant magnetic field. This Hamiltonian without the 
external magnetic field is symmetric under unitary 
transformations, a fact that will determine the critical 
behavior.

The first step to get the partition function is to 
diagonalize the couplings, 

\begin{equation}
\begin{split}
&{\bf \s}_i(\tau)=\sum_\lambda {\bf \s}_{\lambda}(\tau) 
e_i^\lambda, \\
&{\bf \s}_{\lambda}(\tau)=\sum_{i} {\bf 
\s}_i(\tau)e_i^\lambda, 
\label{ortho}
\end{split}
\end{equation}
where $e_i^\lambda$ is the normalized eigenvector of the 
coupling matrix $J_{ij}$. 

Keeping in mind its ill-definedness, we may write the 
partition function sum as a continuum expression,

\begin{equation}
\begin{split}
Z= \int D \tau D \s D \s^\ast \exp \bigg\{ -\int d \tau 
\sum_{a, \lambda} \bigg[ {\s^a_\lambda}^\ast(\tau) \frac{d 
\s^a_\lambda(\tau)}{d \tau} + \mu(\tau) \left( 
{\s^a_\lambda}^\ast(\tau) \s^a_\lambda(\tau -d\tau) - 
Nm\sigma \right) \\
- J_\lambda {\s^a_\lambda}^\ast(\tau) \s^a_\lambda(\tau 
-d\tau) - \frac{1}{\sqrt{2}} \Gamma_\lambda \left( 
{\s^a_\lambda}^\ast(\tau) +  \s^a_\lambda(\tau-d\tau) 
\right) \bigg] \bigg\}
\end{split}
\end{equation}

In discrete notation, the action of eq. (\ref{eq:part1}) 
reads

\begin{equation}
\begin{split}
{\bf A} = \sum_j \epsilon \bigg\{\frac{1}{\epsilon} 
\sum_{a,\lambda} \bigg[\s^{a \,\ast}_{\lambda, j} 
\s^{a}_{\lambda, j} - \s^{a \,\ast}_{\lambda, j} 
\s^{a}_{\lambda, j-1} \bigg] + \mu(j \epsilon) 
\bigg[ \sum_{a, \lambda}  \s^{a \, \ast}_{\lambda,j} 
\s^{a}_{\lambda,j-1} - N m \sigma \bigg] \\
- \sum_{a, \lambda} J_\lambda \s^{a \, 
\ast}_{\lambda, j} \s^{a}_{\lambda, j-1} - 
\sum_{a,\lambda} \Gamma_\lambda 
\frac{(\s^{a \, \ast}_{\lambda, j} + \s^{a}_{\lambda, 
j-1})}{\sqrt{2}} \bigg\}
\end{split}
\end{equation}
where $\epsilon=d \tau$ is the imaginary time step, $j$ 
the time index and $\Gamma_\lambda = \sum_i \Gamma_i 
e^i_\lambda$ is the field in the basis of eigenvectors of 
$J_{ij}$. Collecting all terms we have

\begin{equation}
Z = \int D\mu \prod_{\lambda, a} \left\{ \int \prod_j 
\left(\frac{d\s^{a \, \ast}_{\lambda, j} d\s^{a}_{\lambda, 
j}}{2 \pi i} \right) \exp \left[ - \sum_{ij} \s^{a \, 
\ast}_{\lambda, i} {\bf B_{ij}} \s^{a}_{\lambda, j} + 
\sum_j \epsilon \Gamma_\lambda \frac{(\s^{a \, 
\ast}_{\lambda, j} + \s^{a}_{\lambda, j-1})}{\sqrt{2}} 
\right]\right\} e^{\sum_j N m \sigma \epsilon 
\mu(j\epsilon)}
\end{equation}
where ${\bf B_{ij}} =  {\bf \delta_{ij}} - (1 + \epsilon 
J_\lambda -\epsilon \mu(j \epsilon)) {\bf 
\delta'_{i,j+1}}$; here the prime stands for the fact that 
${\bf \delta'_{1,M+1}}\equiv1$ due to the trace 
structure of the partition function. We can now integrate 
over the spins

\begin{equation}
Z= \int D \mu \exp \left[ \sum_{\lambda,a} \left\{ -m \ln 
\det {\bf B_{ij}} + \frac{\epsilon^2 \Gamma_\lambda^2}{2} 
\sum_{ij} {\bf B_{ij}}^{-1} + m \sigma \epsilon \sum_j \mu 
(j \epsilon) \right\} \right]
\label{eq:compZmiddle}
\end{equation}

As usual, in thermodynamics, one-time quantities like 
$\mu(\tau)$ can be taken independent of $\tau$.
We will employ this simplification throughout the rest of 
this paper.  The determinant and the matrix inversion 
can then be performed \cite{negele}. Integrating over 
$\mu$ by the saddle point method we obtain 

\begin{equation}
\beta F = - m\sigma\beta\mu + \frac{1}{N}\sum_{\lambda,a} 
\left\{ \ln(1 - a_\lambda) - 
\frac{M\epsilon^2\Gamma_\lambda^2}{2(1 - a_\lambda)} 
\right\}
\end{equation}
where $a_\lambda=1 - \epsilon (\mu - J_\lambda)$. Sending 
$M \to \infty$ we finally get

\begin{equation}
\begin{split}
\beta F&=-\beta \mu m \sigma + \frac{m}{N} \sum_\lambda 
\left\{ \ln (1-e^{-\beta (\mu - J_\lambda)}) - 
\frac{\beta \Gamma^2_\lambda}{2(\mu-J_\lambda)} \right\} = 
\\
&= -\beta \mu m (\sigma + \frac{1}{2}) + m \int dJ_\lambda 
\rho(J_\lambda) \left\{\ln \left[2 \sinh 
\left(\frac{\beta}{2} (\mu-J_\lambda) \right) \right] - 
\frac{\beta \Gamma^2_\lambda}{2(\mu-J_\lambda)} \right\}
\label{eq:compF}
\end{split}
\end{equation}
where in the last equality we have assumed that the 
couplings satisfy $\frac{1}{N} \sum_\lambda J_\lambda=0$. 
The saddle point equation reads

\begin{equation}
\sigma +1 = \frac{1}{N} \sum_\lambda \left\{ \frac{1}{1 - 
e^{-\beta (\mu - J_\lambda)}} + 
\frac{\Gamma^2_\lambda}{2(\mu - J_\lambda)^2} \right\} = 
\int dJ_\lambda \rho(J_\lambda) \left\{\frac{1}{1 - 
e^{-\beta (\mu - J_\lambda)}} + 
\frac{\Gamma^2_\lambda}{2(\mu - J_\lambda)^2} \right\}
\label{eq:compsaddle}
\end{equation}

The sums over the different eigenvalues of the coupling 
matrix have been changed into integrals. Each $J_\lambda$ 
has a weight in this integral given by $\rho(J_\lambda)$. 
The actual form for this weight function will depend on 
the type of couplings. A set of weight functions for 
ferromagnets in different cubic lattices can be found in 
Ref. \cite{joyce}, and for spin glasses with long range 
interactions in Refs. 
\cite{nieuwenhuizen-1998_1},\cite{kosterlitz-1976_1}.

In Ref. \cite{nieuwenhuizen-1998_1}, where the spherical 
constraint used was the one in eq. (\ref{eq:sphc1}), the 
matrix ${\bf B}$ was different, namely ${\bf B_{ij}} =  
{(1 + \epsilon \mu(j \epsilon))\bf \delta_{ij}} - (1 + 
\epsilon J_\lambda) {\bf \delta'_{i,j+1}}$. Then eq. 
(\ref{eq:compF}) reads

\begin{eqnarray} \label{Theoconn}
\beta F&=&-\beta \mu m \sigmaN + \frac{m}{N} \sum_\lambda 
\left\{ \ln (e^{\beta \mu}-e^{\beta J_\lambda)}) - 
\frac{\beta
\Gamma^2_\lambda}{2(\mu-J_\lambda)} \right\} = \nonumber\\
&=& -\beta \mu m (\sigmaN - \frac{1}{2}) + m \int 
dJ_\lambda
\rho(J_\lambda) \left\{\ln \left[2 \sinh 
\left(\frac{\beta}{2}
(\mu-J_\lambda) \right) \right] - \frac{\beta
\Gamma^2_\lambda}{2(\mu-J_\lambda)} \right\}
\end{eqnarray}
confirming that the already found shift 
$\sigmaN=\sigma+1$, see eq. (\ref{sisiN=}), indeed brings 
the same value for the free energy. 

At large temperatures these equations reduce to

\begin{equation}
\begin{split}
\beta F= -\beta \mu m \sigmaN + m \int dJ_\lambda 
\rho(J_\lambda) \left\{\ln \beta(\mu-J_\lambda)  - 
\frac{\beta \Gamma^2_\lambda}{2(\mu-J_\lambda)} \right\}
\label{eq:compFTl}
\end{split}
\end{equation}

\begin{equation}
\sigmaN = 
\int dJ_\lambda \rho(J_\lambda) \left\{\frac{T}{\mu - 
J_\lambda} + \frac{\Gamma^2_\lambda}{2(\mu - J_\lambda)^2} 
\right\}
\label{eq:compsaddleTl}
\end{equation}

Apart from a factor two, these are exactly 
the equations of the classical spherical model, see e.g. 
\cite{joyce}. This factor two arises because the momenta 
double the degrees of freedom, see e.g. 
\cite{nieuwenhuizen-1995_2}. Near the phase transition 
they are already approximate, but the transition stays 
within the same universality class.

\subsection{Ferromagnetic couplings with transversal field 
in d dimensions}

In this section we will use the results given in the 
previous one for the concrete case of ferromagnetism 
couplings with uniform transversal field. The Hamiltonian 
in this case differs from the one before eq. 
(\ref{eq:complexham}) in the fact that the couplings only 
act in the $z$-direction while the external field only 
acts in the $x$-direction (we restrict ourselves therefore 
to $m=2$). The free energy reads

\begin{equation}
\beta F=-\beta \mu (2\sigma + 1) + \int \frac{d^d {\bf 
k}}{(2 \pi)^d} \ln \left[2 \sinh \left(\frac{\beta}{2} 
(\mu-J({\bf k})) \right) \right] + 
\ln \left[2 \sinh \left(\frac{\beta \mu}{2} \right) 
\right] - \frac{\beta \Gamma^2}{2\mu} 
\end{equation}
and the saddle point equation

\begin{equation}
2(\sigma +1)= \int \frac{d^d {\bf k}}{(2 \pi)^d} 
\frac{1}{1 - e^{-\beta (\mu - J({\bf k}))}} + \frac{1}{1 - 
e^{-\beta \mu}} + \frac{\Gamma^2}{2 \mu^2}
\label{eq:sadcomp}
\end{equation}
where we have applied the changes $J_\lambda \to J({\bf 
k})$ and

\begin{equation}
\int d J_\lambda \rho(J_\lambda) = \int_{- \pi}^{\pi} 
\frac{d^d {\bf k}}{(2 \pi)^d} 
\end{equation}

We choose $J({\bf k}) \approx J_0 - J' |{\bf k}|^x$
for $|{\bf k}| \to 0$. In the case of short 
range couplings, for instance, one has $x=2$ since $J({\bf 
k})=\sum J\cos k_i\approx J(0)-\half J |{\bf k}|^2$. A 
long range coupling that decays as
$J(r)\sim 1/r^{-\alpha}$ at large $r$ gives $x=\alpha-d$.

As in the theory of Bose-Einstein condensation, 
the saddle point equation fixes the dependence of $\mu$ on 
temperature. There should be a solution at any $T$. In 
order to have a real free energy, $\mu$ cannot be smaller 
than the maximum value for $J({\bf k})$. Therefore, we 
should investigate the convergence of the integral in the 
limit $\mu \to J_0$. If the integral diverges, $\beta$ 
must go to infinity before $\mu$ reaches $J_0$ in order to 
satisfy the saddle point equation, so there exists a $\mu$ 
for all temperatures and no phase transition occurs. If 
the integral converges, however, there will be a range of 
temperatures in which the saddle point as it stands cannot 
hold. This indicates that we have overlooked a macroscopic 
occupation of the ground state, as occurs in Bose-Einstein 
condensation. The relevant integral behaves as 

\begin{equation}
\int_{-\pi}^{\pi} \frac{d^d {\bf k}}{(2 \pi)^d} \frac{1}{1 
- e^{-\beta (J_0 - J({\bf k}))}} \approx 
\frac{\Omega_d}{(2 \pi)^d} \int_{0} dk 
k^{d-1} \frac{1}{1 - e^{-\beta J' k^x}} \propto \int_0 dk 
k^{d-1-x}
\end{equation}
where $\Omega_d$ is the hypersurface of a sphere in d 
dimensions. At ${\bf k} = 0$, this integral converges for 
$d>x$, hence there will be a phase transition for 
dimensions larger than $x$.

At low temperatures, $\mu$ may get stuck at $J_0$ and the 
saddle point equation as it is in eq. (\ref{eq:sadcomp}) 
is no longer valid. This is because,
as in Bose-Einstein condensation 
calculations, the ground state is not properly included in 
the integral. It should be taken out of the sum before 
this one is converted to an integral. 
This causes a change in the free energy by a factor 
$(\mu-J_0) q$ and the saddle point equation becomes

\begin{equation}
2(\sigma +1) =  \int \frac{d^d {\bf k}}{(2 \pi)^d} 
\frac{1}{1 - e^{-\beta (\mu - J({\bf k}))}} + \frac{1}{1 - 
e^{-\beta \mu}} + \frac{\Gamma^2}{2 \mu^2} + q
\label{eq:sadq}
\end{equation}
where $q=\frac{1}{N} \langle \sop_{k=0}^{z \, \dagger} 
\sop^z_{k=0} \rangle$ is the ground state occupation. 
$q$ can be evaluated from the saddle point equation 
$(\mu-J_0) \sqrt{q} = 0$. Thus when $\mu = J_0$ the 
occupation of the ground state can take 
non-zero values that can be determined using eq. 
(\ref{eq:sadq}). Hence the ground state occupation is 
macroscopic in the ordered phase.  

A transversal field will lower the transition 
temperature. Above a certain value $\Gamma_c$, the 
transition does not exist anymore, thus $T=0, 
\Gamma=\Gamma_c$ is a quantum critical point (for a 
complete study over quantum phase transitions, see
e.g. \cite{sachdev}). We will now first study the 
classical critical point, where $\Gamma = 0$.

\subsubsection{Finite temperature phase transition}

For the dimensions where the phase transition exists, the 
critical temperature is found by solving the equation

\begin{equation}
2(\sigma +1)= \int \frac{d^d {\bf k}}{(2 \pi)^d} 
\frac{1}{1 - e^{-\beta_c 
(J_0 - J({\bf k}))}} + \frac{1}{1 - e^{-\beta_c J_0}}
\label{eq:sadcrit}
\end{equation}

The dependence of the chemical potential on the 
temperature near the transition is the first thing needed. 
To get it, we expand the saddle 
point equation around the critical point $T=T_c+\tau$, 
$\mu = J_0 +\delta\mu$. The integral gives, up to first 
order in $\delta\mu$ and $\tau$

\begin{equation}
\int_{-\pi}^{\pi} \frac{d^d {\bf k}}{(2 \pi)^d} \frac{1}{1 
- e^{-\beta (\mu - J({\bf k}))}} \approx \int_{-\pi}^{\pi} 
\frac{d^d {\bf k}}{(2 \pi)^d} \left[ \frac{1}{1 - 
e^{-\beta_c (J_0 - J({\bf k}))}} + \tau \frac{J_0 - J({\bf 
k})}{4 T_c^2 \sinh^2 \left( \frac{J_0 - J({\bf k})}{2 
T_c} \right)} - \delta\mu \frac{1}{4 T_c \sinh^2 \left( 
\frac{J_0 - J({\bf k})}{2 T_c} \right)} \right]
\label{eq:expansion}
\end{equation}

The coefficient of $\delta\mu$ is an integral that 
diverges for $d\leq2x$. This means that for these 
dimensions the leading term in the $\delta\mu$ 
expansion of eq. (\ref{eq:expansion}) has a power smaller 
than one. For dimensions $d>2x$ we will have $\delta\mu 
\propto \tau$ which will lead to the mean field exponents, 
$2x$ is therefore the upper critical dimension. To study 
the system near the critical point we subtract eq. 
(\ref{eq:sadcrit}) from the saddle point equation, 
a procedure that will cancel 
the zeroth order term in the expansion in $\tau$ and 
$\delta\mu$, giving finally

\begin{alignat}{3}
\tau & \approx a_{d<2x} 
\delta\mu^{\frac{d-x}{x}},\quad &a_{d<2x}&=\alpha \frac{4 
\Omega_d T_c^3 \pi}{(2 \pi)^d 
{J'}^{\frac{d}{x}} x \sin \left( \frac{(d-x) \pi}{x} 
\right)} &\text{for } x<d<2x \notag \\
\tau & \approx a_{d=2x} 
\delta\mu \ln \delta\mu,\quad &a_{d=2x}&=\alpha \frac{4 
\Omega_d T_c^3}{(2 \pi)^d {J'}^2 x} &\text{for } d=2x  
\label{eq:dtaucomp} \\
\tau & \approx s_{d>2x} \delta\mu,\quad &a_{d>2x}&=\alpha 
T_c \left[\int \frac{d^d {\bf k}}{(2 \pi)^d} 
\frac{1}{\sinh^2 \left( \frac{J_0 - J({\bf k})}{2 T_c} 
\right)} + 
\frac{1}{\sinh^2 \left( \frac{J_0}{2 T_c} \right)} 
\right] &\text{for } d > 2x. \notag
\end{alignat}
where 
\begin{equation}
\alpha = \left[ \int_{-\pi}^{\pi} \frac{d^d {\bf k}}{(2 
\pi)^d} \frac{J_0 - J({\bf k})}{\sinh^2 \left(\frac{J_0 - 
J({\bf k})}{2 T_c} \right)} + \frac{J_0}{\sinh^2 
\left(\frac{J_0}{2 T_c} \right)} \right]^{-1}
\end{equation}
is a finite, positive number.

The internal energy of the system reads

\begin{equation}
U = -\mu (2 \sigma + 1) + \int_{-\pi}^{\pi} \frac{d^d {\bf 
k}}{(2 \pi)^d} \frac{\mu-J({\bf k})}{2} \coth \left[ 
\frac{\beta(\mu-J({\bf k}))}{2} 
\right] +\frac{\mu}{2} \coth \left(\frac{\beta \mu}{2} 
\right) - \frac{\Gamma^2}{2\mu} 
\end{equation}
 
The specific heat close to the transition from the 
paramagnetic side can be written as

\begin{equation}
C \approx 
\begin{cases}
C_0 +  \frac{x}{a_{d<2x}(d-x)} C_{1}
\tau^{\frac{2x-d}{d-x}} & \text{for } x<d<2x \\
C_0 +  \frac{1}{a_{d>2x}} C_{1}
& \text{for } d>2x
\label{eq:specific}
\end{cases}
\end{equation}
where

\begin{equation}
C_0=\frac{1}{4 T^2} \int_{-\pi}^{\pi} \frac{d^d {\bf 
k}}{(2 \pi)^d} \frac{[\mu - J({\bf k})]^2}{\sinh^2 \left( 
\frac{\mu - J({\bf k})}{2 T} \right)} + \frac{\mu^2}{4 T^2 
\sinh^2 \left( \frac{\mu}{2 T} \right)}
\end{equation}

\begin{equation}
C_{1}= -2\sigma-1 + \int_{-\pi}^{\pi} \frac{d^d {\bf k}}{(2 
\pi)^d} \left\{ \frac{1}{2} \coth \left[ \frac{\mu-J({\bf 
k})}{2T} \right] - \frac{\mu - J({\bf k})}{4 T \sinh^2 
\left( \frac{\mu - J({\bf k})}{2 T} \right)} \right\} + 
\frac{1}{2} \coth \left(\frac{\mu}{2T} 
\right) - \frac{\mu}{4 T \sinh^2 \left( \frac{\mu}{2 T} 
\right)} + \frac{\Gamma^2}{2 \mu^2}
\end{equation}
where $a_{d}$ are the prefactors in eq. (\ref{eq:dtaucomp}) 
for the corresponding dimension. In the ordered phase 
$\mu$ is stuck in its minimum value ($\mu=J_0$) for any 
temperature. Hence, $C = C_0(\mu=J_0)$ in the ordered 
phase. The critical exponent $\alpha$ is the expected one:
$\alpha=\frac{d-2x}{d-x}$ for $x<d<2x$, and the mean field 
value $\alpha=0$ holds for $d>2x$, which describes a jump 
in the specific heat.

Adding a small longitudinal field $h$, the free energy 
reads

\begin{equation}
\beta F = -\beta\mu (2 \sigma+1) + \int \frac{d^d {\bf 
k}}{(2 \pi)^d} \ln \left[2 \sinh \left(\frac{\beta}{2} 
(\mu-J({\bf k})) \right) \right] + 
\ln \left[2 \sinh \left(\frac{\beta \mu}{2} \right) 
\right] - \frac{\beta \Gamma^2}{2\mu} -\frac{\beta 
h^2}{2(\mu - J_0)}
\end{equation}
and the saddle point equation becomes

\begin{equation}
2(\sigma +1)= \int \frac{d^d {\bf k}}{(2 \pi)^d} 
\frac{1}{1 - e^{-\beta (\mu - J({\bf k}))}} + \frac{1}{1 - 
e^{-\beta \mu}} + \frac{\Gamma^2}{2 \mu^2} + 
\frac{h^2}{2(\mu - J_0)^2}
\label{eq:compsada} 
\end{equation}

By differentiating the free energy with respect to $h$ can 
be seen that the magnetization is 
$M_z=\frac{h}{\mu-J_0}$. In the limit $h\to0$, it is 
proportional to 
the square root of the occupation of the ground state, 
since by comparing eq. (\ref{eq:compsada}) 
with eq. (\ref{eq:sadq}) one finds $q=\frac{1}{N} \langle 
\sop_{k=0}^{z \, \dagger} \sop^z_{k=0} \rangle = 
\frac{M^2_z}{2}$. The factor $\frac{1}{2}$ appears because 
is actually the real part of the spin field the one 
macroscopically occupied and a half term appears in the 
change eq. (\ref{eq:cran}). From eq. (\ref{eq:compsada}), 
we can approach the transition by sending 
the longitudinal field to zero at the critical temperature. 
The saddle point equation now accounts for the dependence 
of the chemical potential on the field.
The calculation is similar, yielding finally 

\begin{align}
h & \approx \left(\frac{2 \Omega_d T_c \pi}{(2\pi)^d 
{J'}^{d/x} x 
\sin\left(\frac{\pi(d-x)}{x}\right)} \right)^{\frac{1}{2}} 
\delta\mu^{\frac{d+x}{2x}} & \text{for } x<d<2x \notag \\
h & \approx \left[\frac{2 \Omega_d T_c}{(2\pi)^d {J'}^2 x} 
\ln \delta\mu \right]^{\frac{1}{2}} 
\delta\mu^{\frac{3}{2}} & \text{for } d = 2x \\
h & \approx \left( \int \frac{d^d {\bf k}}{(2 \pi)^d} 
\frac{1}{2 T_c \sinh^2 \left( \frac{J_0 - J({\bf k})}{2 
T_c} \right)} + \frac{1}{2 T_c \sinh^2 \left( \frac{J_0}{2 
T_c} \right)} \right)^{\frac{1}{2}} 
\delta\mu^{3/2} & \text{for } d>2x \notag
\end{align}

Therefore the critical exponent $\delta$ is given by
$\delta = \frac{d+x}{d-x}$ for dimensions $x<d<2x$ and the 
mean field value $\delta=3$ is recovered for $d>2x$. From 
the magnetization, the susceptibility follows as $\chi 
\approx \frac{1}{\delta\mu}$. Therefore we find 
$\gamma=\frac{x}{d-x}$ for $x<d<2x$ and $\gamma=1$ for 
$d>2x$. In the ordered phase, the expansion of the saddle 
point equation, eq. (\ref{eq:compsada}), for $T$ near the 
transition yields

\begin{equation}
M^2_z \approx \tau \frac{1}{2 T_c^2} \left[ 
\int_{-\pi}^{\pi} \frac{d^d {\bf k}}{(2 \pi)^d} \frac{J_0 
- J({\bf k})}{\sinh^2 \left(\frac{J_0 - J({\bf k})}{2 T_c} 
\right)} + \frac{J_0}{\sinh^2 \left(\frac{J_0}{2 T_c} 
\right)} \right] 
\end{equation}

Therefore for all dimensions where the phase transition 
exists, one has $\beta = \frac{1}{2}$.

For other critical exponents the correlation function is 
needed. It can be computed adding the right source term to 
the Hamiltonian, $\sum_\lambda (g_\lambda(\tau_q) {\bf 
\sop}^\dagger_\lambda + g^\ast_{\lambda}(\tau_r) {\bf 
\sop}_\lambda)$ and differentiating

\begin{equation}
G(\lambda , \tau_q | \lambda' ,\tau_r) = < 
\text{T}[\sop^{z \, \dagger}_\lambda 
(\tau_q) \sop^{z}_{\lambda'} (\tau_r)]> = 
\delta_{\lambda,\lambda'} 
\frac{\partial^2}{\partial g^\ast_\lambda(\tau_q) \partial 
g_{\lambda'}(\tau_r)} \frac{Z(g^\ast,g)}{Z_0} 
\bigg\vert_{g^\ast=g=0}
\end{equation}
where T stands for the time ordered product. 
$Z(g^\ast,g)$ is the partition function of the Hamiltonian 
including the source terms and $Z_0$ is the partition 
function without them. This procedure is carefully 
explained in Ref. ~\cite{negele} giving the result

\begin{equation}
G(k,\tau)=G(k,\tau|k,0)= e^{\tau(\mu - J(k))} \left\{ 
\theta(\tau - \eta)(1+n_k) + \theta(-\tau + \eta) n_k 
\right\}
\label{eq:G}
\end{equation}
where $\theta$ is a Heaviside step function and
$\eta$ is a positive infinitesimal that indicates 
that the second term is the relevant one at $\tau=0$. 
Furthermore,

\begin{equation}
n_k = \frac{1}{e^{\beta \omega} -1}
\label{eq:nk}
\end{equation}
is the boson occupation probability, with $\omega=(\mu - 
J({\bf k}))$. Fourier transforming this 
last result to Matsubara frequencies we get

\begin{equation}
G (k, \omega_n) = \frac{1}{J(k) - \mu - i \omega_n} 
\xrightarrow{k\to0} \frac{-1}{J' |k|^x + \delta\mu + i 
\omega_n}
\label{eq:greencomp}
\end{equation}

So when we approach the critical point, we can see from 
this equation that $\xi^{-x} \propto \delta\mu$, then 
using eq. (\ref{eq:dtaucomp}) we find that $\nu = 
\frac{1}{d-x}$ for dimensions $x < d < 2x$ and $\nu = 
\frac{1}{x}$ for $d>2x$. $\eta = 2-x$ due to the fact that 
the couplings depend on $k^x$, and $z=x$ because in the 
denominator $\omega$ appears as a linear term. Both $\eta$ 
and $z$ are valid for any dimension.

This finally gives all the critical exponents of this 
finite temperature phase transition, which are exactly the 
same as in the classical model. This is expected from 
renormalization group arguments \cite{hertz-1976_1}. The 
critical behavior is controlled by a classical 
fixed point, therefore quantum dynamics does not play a 
qualitatively new role. Hence, the results are the same as 
in the classical spherical model \cite{joyce} or 
other models with different quantum dynamics considered at 
finite temperatures \cite{vojta-1996_1}. 

\subsubsection{T=0 quantum phase transition} \label{compT0}

In this section we analyze the behavior of the system at 
$T=0$. As it can be seen from eq. (\ref{eq:sadcomp}), when 
the transversal field increases, the temperature of the 
transition decreases till it reaches zero. This 
defines a quantum critical point $T_c=0$ at 
$\Gamma=\Gamma_c$. In order to study it, an analogous 
procedure as before should be followed. At 
$T=0$ everything happens to be rather simple. The free 
energy reduces to 

\begin{align}
F = - 2 \sigma \mu - \frac{\Gamma^2}{2 \mu}
\label{eq:Fcomp0}
\end{align}

The saddle point equation turns out to be

\begin{align}
2 \sigma = \frac{\Gamma^2}{2 \mu^2} & \qquad \text{in the 
paramagnetic phase,} \notag \\
2 \sigma = \frac{\Gamma^2}{2 J_0^2} + q & \qquad \text{in 
the ferromagnetic phase.}
\label{eq:sadcomp0}
\end{align}
where $q=\half M^2_z$ is the occupation of the ground 
state for small transversal fields. Since the temperature 
vanishes, quantum fluctuations, controlled by $\Gamma$, 
give rise to the phase transition. 
Therefore, the parameter that should be used to control the 
transition is the transversal field and not the 
temperature. Then the proper analog of the specific heat 
will be proportional to the second derivative of the 
free energy with respect to the source of fluctuations, 
the transversal field. 

\begin{equation}
C_\Gamma \equiv\frac{\partial^2 F}{\partial \Gamma^2} = 
-\frac{1}{\mu} + \frac{\Gamma}{\mu^2} \frac{\partial 
\mu}{\partial \Gamma}
\label{eq:Ccomp0}
\end{equation}

As before, we must know the dependence of $\delta\mu$ 
($\mu = J_0 + \delta\mu$) on the distance to the critical 
point ($\delta \Gamma$) in the paramagnetic phase. 
From eqs. (\ref{eq:Ccomp0},\ref{eq:sadcomp0}), it can be 
seen that the analog of the specific heat has a jump 
discontinuity, implying $\alpha=0$. 

\begin{align}
C_\Gamma = 0 & \qquad \text{in the 
paramagnetic phase,} \notag \\
C_\Gamma = \frac{-1}{J_0} & \qquad \text{in the 
ferromagnetic phase.}
\end{align}
where the minus sign comes from the fact that the $T=0$ 
free energy, eq. (\ref{eq:Fcomp0}), is negative.
Adding a small longitudinal field as before, we find the 
critical exponent 
$\delta=3$, since $M_z \propto \frac{h}{\delta\mu}$ and 
$\delta\mu \propto 
h^{\frac{2}{3}}$ at $\Gamma_c$. For the susceptibility, we 
find $\gamma=1$ since 
$\chi \propto \delta\mu^{-1} \propto \delta \Gamma ^{-1}$. 
Subtracting the saddle point equation near the transition 
in the ferromagnetic phase from the one at the transition, 
we get $q=\frac{\Gamma^2 -\Gamma^2_c}{2J_0^2}$, which in 
the lowest order gives $M^2_z\propto \delta 
\Gamma$ and therefore $\beta= \frac{1}{2}$. Eq. 
(\ref{eq:greencomp}) can be used here once it is 
transformed to real frequencies, $i \omega_n = 
\omega + i \eta$. Then since $\delta\mu \propto \delta 
\Gamma$ we find $\nu = \frac{1}{x}$, $\eta = 2 -x$ and 
$z=x$. Hence we find always the mean field values. This 
occurs because the quantum critical point of a 
d-dimensional model shares the critical exponents of a 
classical critical point of a (d+z)-dimensional model, as 
it was shown by general renormalization group arguments 
\cite{hertz-1976_1}. Then, since the 
transition exists for $d>x$, the quantum critical point 
behaves as the classical one for $d>x+z = 2x$. $2x$ is 
just the classical upper critical 
dimension in this model, above which one has the mean 
field values. 

\section{Hamiltonians involving spins but not their 
momenta} \label{realpart}

In this section we are going to extend the analysis of 
section \ref{complex} to a Hamiltonian which only depends 
on the spin operators $\hat S$ and not on the momenta 
$\hat\Pi$. When going from the classical to the quantum 
model, we have to keep in mind that the Hamiltonian must 
be Hermitian. To be precise, the Hamiltonian we will deal 
with is 

\begin{equation}
H = -\frac{1}{2} \sum_{ij} J_{ij} {\bf \hS}_{i} {\bf 
\hS}_{j} - \sum_i \Gamma_i {\bf \hS}_{i}
\label{eq:realham}
\end{equation}
with real valued $J_{ij}$ and $\Gamma_i$ and
where $\hS = (\sop^{\dagger}+\sop)/\sqrt{2}$ is 
the real part of the former spin field. 
Hence, the Hamiltonian does not involve momenta, but the 
spherical constraint does, see eq. (\ref{eq:const2}).
This changes the symmetry of the problem from invariance 
under unitary transformations to orthogonal ones. In terms 
of boson creation and annihilation operators the coupling 
term for symmetric interactions, $J_{ij}=J_{ji}$, is 
proportional to $J_{ij} (2{\bf \sop}_i^\dagger{\bf \sop}_j 
+ {\bf \sop}_i{\bf \sop}_j + {\bf \sop}_i^\dagger{\bf 
\sop}_j^\dagger)$, where we can notice the symmetry of the 
problem. We will see that this model reproduces the 
O($\N$) quantum rotor model. 

We can get the partition function in many ways. A 
similar procedure using discrete imaginary time path 
integrals can be done as before. 
This gives us many problems due to the fact that creation 
and annihilation operators are projected on different time 
steps which is a lengthy and tedious procedure. However, 
the form of the Hamiltonian makes it suitable to apply a 
Bogoliubov transformation (for details see e.g. 
\cite{auerbach}). Due to that, we get the same 
Hamiltonian as before but with different coefficients. In 
order to do so, we must add the spherical constraint 
directly to the Hamiltonian. The procedure is as 
follows: first the couplings matrix is diagonalized by 
inverting the lattice as done before in eq. (\ref{ortho}) 
and then the $\hS$'s are shifted to absorb the field term. 
This finally gives

\begin{equation}
H = \sum_\lambda \left\{ (\mu - \frac{J_\lambda}{2}) {\bf 
\sop}^\dagger_\lambda{\bf \sop}_\lambda - 
\frac{J_\lambda}{4} ({\bf 
\sop}^\dagger_\lambda {\bf \sop}^\dagger_{-\lambda} + {\bf 
\sop}_\lambda {\bf 
\sop}_{-\lambda}) - \frac{\Gamma^2_\lambda}{2(\mu - 
J_\lambda)} \right\} -Nm \mu \sigma
\end{equation}
Performing the Bogoliubov transformation it turns into

\begin{equation}
H = \sum_\lambda \left\{ \sqrt{\mu (\mu - J_\lambda)} 
{\bf \hat \alpha}^{\dagger}_\lambda {\bf \hat 
\alpha}_\lambda - \frac{\Gamma^2_\lambda}{2(\mu - 
J_\lambda)} \right\} -Nm \mu \sigma,
\label{eq:rediag}
\end{equation}
which is a Hamiltonian analogous to eq. 
(\ref{eq:complexham}). So it can 
be diagonalized as explained, giving finally

\begin{equation}
\beta F = -\beta \mu m (\sigma + \frac{1}{2}) + m\int d 
J_{\lambda} \rho(J_{\lambda}) \left\{ \ln \left[ 2 \sinh 
\left( \frac{\beta}{2} \sqrt{\mu ( \mu - J_{\lambda} )} 
\right) \right] - \frac{\beta \Gamma_\lambda^2}{2 (\mu 
-J_\lambda)} \right\} 
\label{eq:Fenergy}
\end{equation}
where we have put back the factor $m$ standing for the 
number of components of the vector spin. The saddle point 
equation is obtained as

\begin{equation}
\sigma +\frac{1}{2} = \int d J_{\lambda} 
\rho(J_{\lambda})\left\{ \frac{2\mu - 
J_{\lambda}}{4 \sqrt{\mu ( \mu - J_{\lambda} )}} \coth 
\frac{\beta}{2} \sqrt{\mu ( \mu - J_{\lambda} )} + 
\frac{\beta \Gamma_\lambda^2}{2 (\mu -J_\lambda)^2} 
\right\}
\label{eq:saddle}
\end{equation}

At large temperatures these equations reduce to

\begin{equation}
\beta F = -\beta \mu m (\sigma + \half)  + m\int d 
J_{\lambda} \rho(J_{\lambda}) \left\{ \ln \beta 
\sqrt{\mu ( \mu - J_{\lambda} )} - \frac{\beta 
\Gamma_\lambda^2}{2 (\mu -J_\lambda)} \right\} 
\label{eq:FenergyTl}
\end{equation}

\begin{equation}
\sigma +\half = \int d J_{\lambda} \rho(J_{\lambda}) 
\left\{  \frac{T}{2\mu} +\frac{T} 
{2( \mu - J_{\lambda} )} + \frac{\beta 
\Gamma_\lambda^2}{2 (\mu -J_\lambda)^2} \right\}
\label{eq:saddleTl}
\end{equation}

These equations are very similar to the standard ones of 
the classical spherical model (up to a factor $2$), see 
eq. (\ref{eq:compFTl}), but they are only identical where 
they should be, namely at large $T$, where also $\mu\sim 
T$ is very large, see also Ref. ~\cite{joyce}. 

\subsection{Ferromagnetic couplings in the presence of a 
transversal field}

Analyzing the phase transition of the Hamiltonian that 
does not depend on the momenta is analogous to the 
previous case.  We begin again by choosing the coupling 
term in the $z$ direction and the external field in the 
$x$ and we assume ferromagnetic couplings. The saddle 
point equation gives a phase transition via a macroscopic 
occupation of the ground state, which in the present case 
is a bit more complicated. The critical exponents are 
different, due to the fact that the symmetries of the 
system have changed. The free energy reads

\begin{equation}
\beta F=-\beta \mu (2\sigma + 1) + \int \frac{d^d {\bf 
k}}{(2 \pi)^d} \ln 
\left[2 \sinh \left(\frac{\beta}{2} \sqrt{\mu (\mu-J({\bf 
k}))} \right) 
\right] + \ln \left[2 \sinh \left(\frac{\beta \mu}{2} 
\right) \right] - \frac{\beta \Gamma^2}{2\mu} 
\end{equation}
and the saddle point equation

\begin{equation}
4 \sigma +2= \int \frac{d^d {\bf k}}{(2 \pi)^d} \frac{2\mu 
- J({\bf k})}{2 \sqrt{\mu (\mu-J({\bf k}))}} \coth \left[ 
\frac{\beta}{2} \sqrt{\mu (\mu-J({\bf k}))} \right] + 
\coth \left(\frac{\beta \mu}{2} \right) + 
\frac{\Gamma^2}{\mu^2}
\end{equation}

We now analyze this model in detail.

\subsubsection{Finite temperature phase transition}

Following the same procedure as before we can find that 
the transition exists for $d>x$ and that the upper 
critical dimension is $d=2x$. The critical 
temperature is the solution of 

\begin{equation}
4 \sigma +2= \int \frac{d^d {\bf k}}{(2 \pi)^d} 
\frac{2J_0  - J({\bf k})}{2 \sqrt{J_0 (J_0-J({\bf k}))}} 
\coth \left[ \frac{\beta}{2} 
\sqrt{J_0 (J_0-J({\bf k}))} \right] + \coth 
\left(\frac{\beta J_0}{2} \right)
\end{equation} 

The dependence of the chemical potential in the 
temperature near the classical critical point reads

\begin{gather}
\begin{align}
\tau & \approx a_{d<2x} \delta\mu^{\frac{d-x}{x}},\quad 
&a_{d<2x}&=\alpha \frac{2 \Omega_d T_c^3 \pi}{(2 \pi)^d 
{J'}^{\frac{d}{x}} x \sin \left( \frac{(d-x) \pi}{x} 
\right)} &\text{for } x<d<2x  \notag \\
\tau & \approx a_{d=2x}
\delta\mu \ln \delta\mu,\quad &a_{d=2x}&=\alpha \frac{2 
\Omega_d T_c^3}{(2 \pi)^d {J'}^2 x} &\text{for } d=2x  
\\
\tau & \approx  a_{d>2x}\delta\mu,\quad &&  &\text{for } 
d > 2x  \notag
\end{align}\\
a_{d>2x}=\alpha \left( \frac{\partial}{\partial \mu} 
\left\{\int \frac{d^d {\bf k}}{(2 \pi)^d} \frac{2\mu - 
J({\bf k})}{2 \sqrt{\mu (\mu-J({\bf k}))}} \coth \left[ 
\frac{\beta}{2} \sqrt{\mu (\mu-J({\bf k}))} \right] + 
\coth \left(\frac{\beta \mu}{2} \right) \right\} 
\right)_{\mu=J_0} \notag
\label{eq:retd}
\end{gather}
where 
\begin{equation}
\alpha = \left[ \int_{-\pi}^{\pi} \frac{d^d {\bf k}}{(2 
\pi)^d} \frac{2 J_0 - J({\bf k})}{2 \sinh^2 
\left(\frac{\sqrt{J_0(J_0 - J({\bf k}))}}{2 T_c} \right)} 
+ \frac{J_0}{\sinh^2 \left(\frac{J_0}{2 T_c} \right)} 
\right]^{-1}
\end{equation}

The internal energy of the system reads

\begin{equation}
U = -\mu (2 \sigma + 1) + \int_{-\pi}^{\pi} \frac{d^d {\bf 
k}}{(2 \pi)^d} \frac{\sqrt{\mu (\mu-J({\bf k}))}}{2} \coth 
\left[ \frac{\beta}{2} \sqrt{\mu (\mu-J({\bf k}))} \right] 
+ \frac{\mu}{2}\coth \left(\frac{\beta \mu}{2} \right) - 
\frac{\Gamma^2}{2 \mu}
\end{equation}

The specific heat has the same expression as in eq. 
(\ref{eq:specific}) where $a_d$ now correspond to the 
prefactors of eq. (\ref{eq:retd}) and with coefficients

\begin{equation}
C_0=\frac{1}{4 T^2} \int_{-\pi}^{\pi} \frac{d^d {\bf 
k}}{(2 \pi)^d} 
\frac{\mu (\mu - J({\bf k}))}{\sinh^2 \left( 
\frac{\sqrt{\mu (\mu-J({\bf 
k}))}}{2 T} \right)} + \frac{\mu^2}{4 T^2 \sinh^2 \left( 
\frac{\mu}{2 T} \right)}
\end{equation}

\begin{equation}
\begin{split}
C_{1}=& -(2\sigma+1) + \int_{-\pi}^{\pi} \frac{d^d {\bf 
k}}{(2 \pi)^d} \left\{ \frac{2 \mu - J({\bf k})}{4 
\sqrt{\mu (\mu-J({\bf k}))}} 
\coth \left( \frac{\beta}{2} \sqrt{\mu (\mu-J({\bf k}))} 
\right) - \frac{2\mu - J({\bf k})}{8 T \sinh^2 \left( 
\frac{\sqrt{\mu(\mu - J({\bf k}))}}{2 T} \right)} \right\} 
\\
&+ \frac{1}{2} \coth \left(\frac{\mu}{2T} \right) - 
\frac{\mu}{4 T \sinh^2 \left( \frac{\mu}{2 T} \right)}
\end{split}
\end{equation}

This is analogous to the previous model and gives the same 
exponent, $\alpha=\frac{d-2x}{d-x}$ for $x<d<2x$ and 
$\alpha=0$ for $d>2x$. Adding a small magnetic field 
longitudinal to the couplings, the free energy becomes

\begin{equation}
\beta F=-\beta \mu (2\sigma + 1) + \int \frac{d^d {\bf 
k}}{(2 \pi)^d} \ln \left[2 \sinh \left(\frac{\beta}{2} 
\sqrt{\mu (\mu-J({\bf k}))} \right) 
\right] + \ln \left[2 \sinh \left(\frac{\beta \mu}{2} 
\right) \right] - \frac{\beta \Gamma^2}{2\mu} - 
\frac{\beta h^2}{2 (\mu - J_0)}
\end{equation}
therefore the magnetization is $M_z=\frac{h}{\mu - J_0}$, 
which is as before the square root of the occupation of 
the ground state $q=\frac{1}{N} < \hS_z(|{\bf 
k}|=0)^2>=M^2_z$. The saddle point equation is now

\begin{equation}
2 (2 \sigma + 1) = \int_{-\pi}^{\pi} \frac{d^d {\bf k}}{(2 
\pi)^d} \frac{2 \mu - J({\bf k})}{2 \sqrt{\mu (\mu - 
J({\bf k}))}} \coth \left\{ \frac{\beta}{2} \sqrt{ \mu 
(\mu - J({\bf k}))} \right\} + \coth \left( 
\frac{\beta \mu}{2} \right) + \frac{\Gamma^2}{\mu^2} + 
\frac{h^2}{(\mu - J_0)^2}
\end{equation}

With all these and following the algebra of the previous 
section one finds the same critical exponents for the 
magnetization for the same dimensions since we are in the 
classical critical point.

The time ordered correlation function 
$\langle$T$\hS^z_k(\tau)\hS^z_{-k}(0)\rangle$ differs from 
the previous one, eq. (\ref{eq:G}) since in this case the 
$\hS$ are not the variables that diagonalize the 
Hamiltonian in eq. (\ref{eq:rediag}). We must write $\hS$ 
in terms of $\bf \hat \alpha$ and then compute the 
correlations. This brings

\BEQ
G(k,\tau|k,0)=\frac{J({\bf k})}{4\sqrt{\mu(\mu-J({\bf 
k}))}} \left\{n_k \cosh \left[\tau \sqrt{\mu(\mu-J({\bf 
k}))}\right] + e^{-|\tau| \sqrt{\mu(\mu-J({\bf k}))}} 
\right\}
\EEQ
where 

\BEQ
n_k=\frac{1}{e^{\beta \sqrt{\mu(\mu-J({\bf k}))}}-1}
\EEQ
which in frequency space reads

\BEQ
G(k,i \omega_n) = \frac{-J({\bf k})}{2[\omega_n^2 - 
\mu(\mu-J({\bf k}))]}
\EEQ

When approaching the critical point we find that $\xi^{-x} 
\propto \delta\mu$ as before and we get the same value 
$\nu=\frac{1}{d-x}$ for dimensions $x<d<2x$ and 
$\nu=\frac{1}{x}$ for $d>2x$. Since couplings appear in 
the same way as before we also get the same value, 
$\eta=2-x$ for all dimensions.. The difference appears in 
the dynamical critical exponent. Here $\omega_n$ appears 
squared, therefore $z=x/2$. Here we see how the model 
reproduces the critical exponents of the rotor model as in 
Ref. \cite{vojta-1996_1} bringing thus a different 
behavior at the $T=0$ quantum critical point from the 
model of section \ref{compT0}

\subsubsection{T=0 quantum phase transition}

In this case the $T=0$ phase transition is more 
interesting due to the fact that the dynamical critical 
exponent $z=\frac{x}{2}$ is smaller than $z=x$ of the 
previous section. The free energy reads

\BEQ
F=-\mu(2\sigma +1)+ \int \frac{\d^d {\bf k}}{(2\pi)^d} 
\frac{\sqrt{\mu(\mu-J({\bf k}))}}{2} + \frac{\mu}{2} - 
\frac{\Gamma^2}{2\mu}
\EEQ
and the saddle point is set by

\BEQ
4\sigma +1 = \int \frac{\d^d {\bf k}}{(2\pi)^d} 
\frac{2\mu-J({\bf k})}{2\sqrt{\mu(\mu-J({\bf k}))}} + 
\frac{\Gamma^2}{\mu^2}
\EEQ

We find that the transition exists for dimensions larger 
than $d>\frac{x}{2}$ and $d=\frac{3x}{2}$ is the upper 
critical dimension. The chemical potential depends on the 
source of fluctuations $\delta\Gamma= \Gamma-\Gamma_c$ 
as

\begin{gather}
\begin{align}
\delta\Gamma & \approx a_{d<3x/2} 
\delta\mu^\frac{2d-x}{2x},\;  &a_{d<3x/2}&=\frac{\Omega_d 
J_0^{\frac{5}{2}}}{2 (2\pi)^d {J'}^{\frac{d}{x}}} 
\frac{\Gamma\left(\frac{3}{2}-\frac{d}{x}\right)\Gamma
\left(\frac{d}{x}\right)}{(2d-x)x\sqrt{\pi}} & \text{for } 
\frac{x}{2} < d < \frac{3x}{2}\notag \\
\delta\Gamma & \approx a_{d=3x/2}\delta\mu \ln 
\delta\mu,\; &a_{d=3x/2}&=\frac{\Omega_d}{(2\pi)^d}
\frac{J_0^{\frac{5}{2}}}{8x{J'}^{\frac{3}{2}}}& \text{for 
} d = \frac{3x}{2} \\
\delta\Gamma & \approx a_{d>3x/2} \delta\mu,\;  & &  & 
\text{for } d>\frac{3x}{2} \notag
\end{align}\\
a_{d=3x/2}=\frac{J_0^2}{2}\left[\frac{2\Gamma_c^2}{J_0^3}
-\int\frac{\d^d {\bf k}}{(2\pi)^d} \left\{ 
\frac{1}{\sqrt{J_0(J_0-J({\bf k}))}} - \frac{2J_0-J({\bf 
k})}{4\left(J_0(J_0-J({\bf k}))\right)^{\frac{3}{2}}} 
\right\} \right] \notag
\label{eq:gammadelta}
\end{gather}
where the $\Gamma$'s on the right hand side of the first 
equality are Euler's Gamma functions.
The specific heat (see eq. (\ref{eq:Ccomp0}) ) coming from 
the disordered region will behave as

\begin{equation}
C_\Gamma \approx 
\begin{cases}
\frac{-1}{J_0} - \frac{4 \Gamma_c}{a_{d<\frac{3x}{2}} 
J_0^2 (2d-x)} \delta\Gamma^{\frac{-2d+3x}{2d-x}} & 
\text{for } \frac{x}{2}<d<\frac{3x}{2} \\
\frac{-1}{J_0} + \frac{2 \Gamma_c}{J_0^4} 
\left(\frac{2\Gamma_c^2}{J_0^3} - a_{d>\frac{3x}{2}} 
\right)^{-1} & \text{for } d>\frac{3x}{2} 
\end{cases}
\end{equation}
where $a_d$ is the prefactor in eq. (\ref{eq:gammadelta}) 
for the proper dimension. Coming from the ordered region, 
conversely, $C_\Gamma \approx \frac{-1}{J_0}$. Therefore 
$\alpha=\frac{2d-3x}{2d-x}$ for 
$\frac{x}{2}<d<\frac{3x}{2}$ and $\alpha=0$ for 
$d>\frac{3x}{2}$.

The dependence of a small longitudinal field on the 
chemical potential, in case the transversal field is at 
its critical value, reads

\begin{align}
h &\approx a_{d<\frac{3x}{2}}^{\frac{1}{2}} 
\delta\mu^{\frac{2d+3x}{4x}} & & \text{for } \frac{x}{2} < 
d < \frac{3x}{2}\notag \\
h & \approx \left[ a_{d=\frac{3x}{2}} \ln 
\delta\mu\right]^{\frac{1}{2}} \delta\mu^{\frac{3}{2}}& 
&\text{for } d = \frac{3x}{2}\\
h & \approx \left(-\frac{2\Gamma_c^2}{J_0^3}-a_{d>
\frac{3x}{2}}\right)^{\frac{1}{2}} \delta\mu^{\frac{3}{2}} 
& & \text{for } d>\frac{3x}{2} \notag
\end{align}

From these equations and the ones for the magnetization 
and the susceptibility we can find that 
$\delta=\frac{2d+3x}{2d-x}$ and $\gamma=\frac{2x}{2d-x}$ 
for $\frac{x}{2} < d < \frac{3x}{2}$, while $\delta=3$ and 
$\gamma=1$ for $d>\frac{3x}{2}$. As before 
$\beta=\frac{1}{2}$ for every dimension. For the 
correlation function the calculation is the same as in the 
finite temperature case, projected into real time, the 
exponents are $\nu=\frac{2}{2d-x}$ for dimensions 
$\frac{x}{2} < d < \frac{3x}{2}$ and $\nu=\frac{1}{x}$ 
above the critical dimension, $\eta=2-x$ and 
$z=\frac{x}{2}$ for all dimensions.

\section{Generalization and mapping from Heisenberg spins} 
\label{mapping}

In this section we generalize the two preceding 
Hamiltonians and we map the Heisenberg model onto the 
spherical model.  In a more compact way, we can write the 
former Hamiltonians in absence of external field as 

\begin{equation}
H = - \sum_{ij} \left( A_{ij} \sop^\dagger_{i} \sop_{j } + 
\frac{B_{ij}}{2} \left[\sop^\dagger_{i} \sop^\dagger_{j} + 
\sop_{i} \sop_{j} \right] \right)
\label{eq:generham}
\end{equation}

If the matrices $A_{ij}$ and $B_{ij}$ can be diagonalized 
simultaneously, the techniques from previous sections can 
be used. The free energy reads 

\begin{equation}
\beta F = -\beta \mu m (\sigma + \frac{1}{2}) + 
\frac{m}{N} \sum_\lambda \left[ \frac{\beta A_\lambda}{2} 
+ \ln \left\{2 \sinh \left(\frac{\beta}{2} \sqrt{( \mu - 
A_\lambda)^2 - B_\lambda^2} \right) 
\right\} \right]
\label{eq:fgener}
\end{equation}
where $\mu$ satisfies the saddle point equation

\begin{equation}
\sigma + \frac{1}{2} = \frac{1}{N} \sum_\lambda \frac{ \mu 
- A_\lambda}{2 \sqrt{( \mu - A_\lambda)^2 - B_\lambda^2}} 
\coth \left\{\frac{\beta}{2} \sqrt{( \mu - A_\lambda)^2 - 
B_\lambda^2} \right\}
\label{eq:sgener}
\end{equation}

The coefficient $B_{ij}$ in eq. (\ref{eq:generham}) is the 
responsible for a change in the symmetries of the problem. 
If $B_{ij}$ is zero, the action is symmetric under unitary 
transformations while if it is non-zero the symmetry is 
reduced to orthogonal. 

The mapping from Heisenberg spins comes as follows. The 
Hamiltonian can be written in terms of Schwinger bosons 
\cite{auerbach}. The Schwinger boson transformation for 
SU($2$) spins reads

\begin{equation}
S^+ = a_1^\dagger a_2, \qquad S^- = a_1a_2^\dagger, \qquad 
S^z = \frac{1}{2}(a_1^\dagger a_1-a_2^\dagger a_2)
\end{equation}

This can be generalized to SU($\N$) spins and expand 
around the large-$\N$ limit \cite{arovas-1988_1}. In a 
path integral formalism for ferromagnetic interactions

\begin{equation}
H = - \frac{1}{2} \sum_{ij} J_{ij} {\bf S_i} \cdot {\bf 
S_j} \to -\frac{1}{2N} \sum_{ij,mn} J_{ij} a^\ast_{jm} 
a_{im} a^\ast_{in} a_{jn} 
\end{equation}
where $i,j$ represent lattice sites and $m,n$ represent 
the boson flavor. The Hilbert space spanned by Schwinger 
bosons is much larger than the one given by Heisenberg 
spins. The constraint needed to restrict it to the physical Hilbert space is that the number of Schwinger 
bosons at each site has to be kept fixed $\sum_m^\N n_m = 
\N S$. This is inserted into the formalism in the same way 
as we have done it with the spherical constraint, a 
Lagrange multiplier $\mu_i(\tau)$ appears. 
The biquadratic terms can be decoupled by a 
Hubbard-Stratonovich transformation (see e.g. 
\cite{negele}). In the case of a ferromagnet, the 
transformation at each time step and for each flavor in 
the path integral reads
 
\begin{equation}
\exp \left\{-\frac{\epsilon}{2N} \sum_{ij} J_{ij} 
a^\ast_{j} a_{i} a^\ast_{i} a_{j} \right\} \propto \int 
\prod_{i,j} dQ_{ij} \exp \left\{\frac{\epsilon \N}{2} 
\sum_{i,j} Q_{ij} J_{ij} Q_{ji} - \frac{\epsilon}{2} 
\sum_{ij} Q_{ij} J_{ij} a^\ast_j a_i \right\}
\end{equation}
where a field $Q_{ij}(\tau)$ has been generated. In the 
mean field approximation, one puts $Q_{ij}(\tau) = Q$ and 
$\mu_i(\tau) = \mu$. Hence, one gets up to a non 
interesting constant

\begin{equation}
H^{FM-B}_{MF} (\N) = \sum_{i,m} \mu a_{im}^\dagger a_{im} 
- Q \sum_{ij,m} J_{ij} a_{jm}^\dagger a_{im} + 
\frac{NQ^2}{2} \sum_{ij} J_{ij} - \N N S \mu
\end{equation}
where we have already added the Schwinger boson 
constraint. The free energy per particle reads

\begin{equation}
\beta F = \frac{\N}{N} \sum_{\bf k} \ln (1 - e^{-\beta(\mu 
- Q J({\bf k})}) + \frac{\beta \N Q^2}{2} J({\bf k}=0) - 
\beta S \mu \N
\end{equation}
and the saddle point equations are

\begin{align}
\frac{1}{N} \sum_{\bf k} n_{\bf k} &= S \\
\frac{1}{N} \sum_{\bf k} J({\bf k}) n_{\bf k} &= Q J({\bf 
k}=0)
\end{align}
where $n_k$ is the boson occupation number eq. 
(\ref{eq:nk}) with $\omega=\mu-QJ({\bf k})$. Subtracting 
the two saddle point equations we can see that for large 
$S$ and small $T$ we can approximate $Q \approx S$ 
recovering then the spherical model 
eq. (\ref{eq:compF},\ref{eq:compsaddle}) for zero external 
field or eq. (\ref{eq:fgener},\ref{eq:sgener}) for 
$B_{ij}=0$. From this approach we thus see that the free 
energy of a SU($\N$) Heisenberg ferromagnet for large $\N$ 
is formally the same as the quantum spherical model 
proposed in eq. (\ref{eq:complexham}) in the thermodynamic 
limit, so when the radius of the hypersphere that defines 
the model ($N$ in eq. (\ref{eq:const2}) ) is also very 
large. Thus the large $\N$ limit is somehow analogous to 
Staley's large spin dimensionality limit.

In the case of an SU($\N$) antiferromagnet the procedure 
is more or less the same but the symmetries are different.
The lattice is divided in two sublattices $A,B$. In one of 
the sublattices a spin rotation is performed that allows 
us to write the Hamiltonian in the form \cite{arovas-1988_1}

\begin{equation}
H = \frac{1}{2} \sum_{ij} J_{ij} S_i \cdot S_j \to 
-\frac{1}{2N} \sum_{ij,mn} J_{ij} a^\ast_{im} a^\ast_{im} 
a_{jn} a_{jn} 
\end{equation}

Performing a Hubbard-Stratonovich transformation as 
before, the Hamiltonian with the Schwinger boson 
constraint in the mean field approximation finally reads

\begin{equation}
H^{AFM-B}_{MF} (\N) = \sum_{i,m} \mu a_{im}^\dagger a_{im} 
- \frac{Q}{2} \sum_{ij,m} J_{ij} (a_{im}^\dagger 
a^\dagger_{jm} + a_{im} a_{jm}) + \frac{NQ^2}{2} \sum_{ij} 
J_{ij} - \N N S \mu
\end{equation}

It is important to stress that here the SU($\N$) symmetry 
has been reduced to a residual O($\N$). The free energy 
per particle reads

\begin{equation}
\beta F = \frac{\N}{N} \sum_{\bf k} \ln \left(2 \sinh 
\left[\frac{\beta}{2} \sqrt{\mu^2 - Q^2 J^2({\bf k})} 
\right] \right) - \beta \N \left( S + \frac{1}{2} 
\right)\mu + \frac{\beta \N Q^2}{2} J({\bf k}=0)
\label{eq:fanti}
\end{equation}
and the saddle point equations read read

\begin{align}
\frac{1}{N} \sum_{\bf k} \frac{\mu}{\sqrt{\mu^2 - Q^2 
J^2({\bf k})}} \left( n_{\bf k} + \frac{1}{2} \right) &= S 
+ \frac{1}{2} 
\label{eq:santi}\\
\frac{1}{N} \sum_{\bf k} \frac{J^2 ({\bf k}) Q 
}{\sqrt{\mu^2 - 
Q^2 J^2({\bf k})}} \left( n_{\bf k} + \frac{1}{2} \right) 
&= Q J({\bf k}=0)
\end{align}
where $n_k$ is eq. (\ref{eq:nk}) for 
$\omega=\sqrt{\mu^2-Q^2J^2({\bf k})}$.
Subtracting the first equation times $\mu$ from the second 
times $Q$ we get

\begin{equation}
\frac{1}{N} \sum_{\bf k} \sqrt{\mu^2 - Q^2 J^2({\bf k})}  
\left( n_{\bf k} - \frac{1}{2} \right) = \mu \left( S + 
\frac{1}{2} \right) -  Q^2 J({\bf k}=0)
\end{equation}

The first term is proportional to $T$, so for very small 
temperatures and very large $S$, near the transition where $\mu \approx Q J({\bf k}=0)$, 
we can approximate $Q \approx S + \frac{1}{2}$. Then eqs. 
(\ref{eq:fanti},\ref{eq:santi}) are analogous to eqs. 
(\ref{eq:fgener},\ref{eq:sgener}) for $A_{ij}=0$. This 
will have the same critical behavior as the model in 
section \ref{realpart} due to the fact that it comes from 
the term $\coth[\sqrt{\mu-J({\bf k})}]/\sqrt{\mu-J({\bf 
k})}$ which also appears here due to the equality $2 
n_{\bf k} + 1=\coth[\sqrt{(\mu + Q J({\bf k}))(\mu - Q 
J({\bf k}))}]$.

\section{Conclusion}

In this paper we have explained a way of working with 
quantum spherical spin models using path integrals and 
coherent states. Some examples of the use of this 
formalism are given, eqs. 
(\ref{eq:complexham},\ref{eq:realham}), and their critical 
phenomena are studied.  We propose a comparison with 
SU($\N$) Heisenberg models that gives a geometrical 
interpretation to the quantum spherical spins. The 
spherical constraint we use, fixes the number of spin 
quanta $\hSig$, eq. (\ref{eq:const2}); in other words, it 
fixes both the average length square of the spin operator, 
$\hS^2$, and the one of its conjugate momentum, $\hPi^2$. 
The usual version of the quantum spherical model, on the 
contrary, involves only the spin part $\hS$. The presence 
of momenta in the spherical constraint allows the 
Hamiltonian to have no kinetic term, since it can be 
induced by the constraint, a fact that can change the 
symmetries of the problem, and due to that, the critical 
behavior.

The Hamiltonian in eq. (\ref{eq:complexham}) yields an 
action invariant under unitary transformations. It brings 
formally the same free energy as a SU($\N$) Heisenberg 
ferromagnet in the limit of large $\N$. The other 
Hamiltonian studied, eq. (\ref{eq:realham}), brings an 
action invariant under orthogonal transformations; it 
gives the same critical behavior as an SU($\N$) Heisenberg 
antiferromagnet in the limit of large $\N$, which is, in 
its turn, analogous to an O($\N$) nonlinear $\sigma$-model 
or quantum rotor model \cite{vojta-1996_1,sachdev}.
The main difference between these models lies
in the dynamical critical exponent $z$ which brings a 
different behavior at the quantum critical point. 
Classical critical phenomena are, as expected, the same in 
both models and equal to those of the classical spherical 
model.

In the formulation of the model, the strict spherical 
constraint has been used where fluctuations on the 
particle number are not allowed. The constraint is added 
to the action via a Lagrange multiplier. The strict 
approach has to be abandoned when we integrate this 
Lagrange multiplier using the saddle point approximation. 
In this step, we automatically allow fluctuations on the 
particle number and therefore the constraint ends being 
satisfied only in average. These effects are immaterial in 
the considered thermodynamic limit, but do enter finite 
size corrections.

The analogy of the two Hamiltonians studied here with 
Heisenberg models in the large spin dimensionality limit 
has a drawback. Both models have different coupling to the 
external field. In spherical models it comes in linearly, 
as a source term. No analog has been found for 
this in the large spin dimensionality limit of the 
Heisenberg model where each spin contribution brings a 
bilinear term in Schwinger bosons. 

Another approach could have been to start directly from 
the SU($\N$) Heisenberg model and to do the already stated 
large $\N$ limit to get to a solvable model. In order to 
have a transversal field that competes with the ordering 
of the interacting spins one could introduce anisotropy in 
the model. A study of this type has been done for 2 
dimensions by Timm et al. \cite{timm-2000_1} in terms of 
Schwinger bosons and in terms of Holstein-Primakoff bosons 
by Kaganov et al. \cite{kaganov-1987_1} for any dimension. 
The anisotropy term brings a residual spin symmetry 
describing Ising or XY spins. The phase transition depends 
on the type of this residual symmetry; an additional 
transversal field decreases the transition temperature 
towards zero giving a quantum critical point, result 
qualitatively reproduced by our model.

In spite of the lack of direct interpretation of the 
source term in the mapping from Heisenberg spins, the 
phase diagram follows the expected behavior for a spin 
model with an external transversal field. The critical 
exponents for the classical and the quantum model are the 
ones expected by renormalization group arguments. The 
quantum critical point behaves as the classical one for 
dimensions $D_{quant}=d_{class} + z$ where $z$ is the 
dynamical critical exponent. For instance, in the model 
with unitary symmetry eq. (\ref{eq:complexham}), $z$ is 
just the difference between the lower and the upper 
critical dimension, thus the quantum critical point has 
the same mean field exponents as the classical finite 
temperature critical point.

Many other models have a clear analog with this one. 
Sachdev and Bhatt \cite{sachdev-1990_1} represented pairs 
of spins in a square lattice with a bond representation; 
they form either a singlet or a triplet. These elements 
can be written down in terms of the canonical ``Schwinger 
boson'' representation of the generators of $SU(2)\otimes 
SU(2)=SO(4)$. Since a couple of spins either form a 
singlet or a triplet, a constraint must be added 
$s^\dagger s + \sum_\alpha t^\dagger_\alpha t_\alpha = 1$, 
where $s$ represents the singlet annihilation operator, 
and $t_\alpha$ represents a triplet annihilation operator 
in the $\alpha$ direction. Sachdev and Bhatt study using 
this formalism systems with interactions up to third 
nearest neighbors. They make the further assumption that 
the singlet part condenses and replace the $s$ operator by 
its mean field value $\langle s \rangle= \bar s$, and 
solve the rest for the triplets. The final Hamiltonian is 
very close to our eq. (\ref{eq:realham}), or, better, the 
generalization of our model eq. (\ref{eq:generham}) with 
the proper couplings. A minor difference is the role 
played by the non-constant mean value of the singlet part.

An interesting line for future research would be to expand 
this version of quantum spherical spin models to different 
types of interactions and fields. Randomness is easily 
added in the model. The dynamics could also be studied for 
the constraint described here. Another line would be to 
study in this approximation the Heisenberg SU($\N$) model 
for any dimension with anisotropy in one dimension and 
transversal field, giving special attention to critical 
phenomena and compare it to the Holstein-Primakoff 
approximation of Ref. ~\cite{kaganov-1987_1}.

\begin{acknowledgments}
The authors would like to thank S.Sachdev for his 
invaluable help and advice. One of us, RSG, would like to 
thank also S.Peysson, J.Zaanen and E.Altman for fruitful 
discussions. This work is part of the research program of 
the 'Stichting voor Fundamenteel Onderzoek der Materie 
(FOM)', which is financially supported by the 'Nederlandse 
Organisatie voor Wetenschappelijk Onderzoek (NWO)'.
\end{acknowledgments}

\bibliography{biblio2}

\end{document}